\documentclass[a4paper,11pt]{article}
\usepackage{jheppub}
\pdfoutput=1
\usepackage{xcolor}
\usepackage{amsmath}
\usepackage{graphicx}
\usepackage{color}
\usepackage{subfigure}
\usepackage{multirow}
\usepackage{makecell}
\usepackage{array}
\definecolor{darkgreen}{rgb}{0,0.5,0}

\newcommand{\rowheight}[1]{\renewcommand{\arraystretch}{#1}}

\title{Three-dimensional Imaging of Pion using Lattice QCD: Generalized Parton Distributions}

\author[a]{Heng-Tong Ding,}  
\author[b]{Xiang Gao,}  
\author[b]{Swagato Mukherjee,}  
\author[b]{Peter Petreczky,}
\author[a,b,c,*]{Qi Shi}

\footnotetext{*Corresponding author.}

\author[d]{Sergey Syritsyn}  
\author[e]{and Yong Zhao}  

\affiliation[a]{Key Laboratory of Quark \& Lepton Physics (MOE) and Institute of Particle Physics, Central China Normal University, Wuhan, Hubei 430079, China}  
\affiliation[b]{Physics Department, Brookhaven National Laboratory, Upton, New York 11973, U.S.A.}  
\affiliation[c]{RIKEN-BNL Research Center, Brookhaven National Laboratory, Upton, New York 11973, U.S.A.}  
\affiliation[d]{Department of Physics and Astronomy, Stony Brook University, Stony Brook, New York 11790, U.S.A.}  
\affiliation[e]{Physics Division, Argonne National Laboratory, Lemont, Illinois 60439, U.S.A.}  

\emailAdd{hengtong.ding@ccnu.edu.cn}  
\emailAdd{xgao@bnl.gov}  
\emailAdd{swagato@bnl.gov}  
\emailAdd{petreczk@bnl.gov}  
\emailAdd{qshi1@bnl.gov}  
\emailAdd{sergey.syritsyn@stonybrook.edu}  
\emailAdd{yong.zhao@anl.gov}

\abstract{
In this work, we report a lattice calculation of $x$-dependent valence pion generalized parton distributions (GPDs) at zero skewness with multiple values of the momentum transfer $-t$. The calculations are based on an $N_f=2+1$ gauge ensemble of highly improved staggered quarks with Wilson-Clover valence fermion. The lattice spacing is 0.04 fm, and the pion valence mass is tuned to be 300 MeV. We determine the Lorentz-invariant amplitudes of the quasi-GPD matrix elements for both symmetric and asymmetric momenta transfers with similar values and show the equivalence of both frames. Then, focusing on the asymmetric frame, we utilize a hybrid scheme to renormalize the quasi-GPD matrix elements obtained from the lattice calculations. After the Fourier transforms, the quasi-GPDs are then matched to the light-cone GPDs within the framework of large momentum effective theory with improved matching, including the next-to-next-to-leading order perturbative corrections, and leading renormalon and renormalization group resummations. We also present the 3-dimensional image of the pion in impact-parameter space through the Fourier transform of the momentum transfer $-t$.}

\arxivnumber{2407.03516}

\begin{document}
\maketitle
\flushbottom

\section{\label{sec: intro}Introduction}

Since its discovery in 1947, the pion has been the subject of intense research, recognized for its dual identity as both a Goldstone boson linked to chiral symmetry breaking and a bound state in quantum chromodynamics (QCD). Over the years, substantial efforts have been dedicated to studying its internal structure by analyzing experimental data. Key methodologies have involved extracting form factors (FFs) through the pion-electron scattering~\cite{NA7:1986vav} and discerning parton distribution functions (PDFs) via the Drell-Yan process~\cite{Conway:1989fs}. However, these approaches are limited to revealing only the one-dimensional structure of the hadron. For a more comprehensive, three-dimensional perspective, the focus has shifted to generalized parton distributions (GPDs), a concept introduced in the 1990s~\cite{Muller:1994ses, Ji:1996ek, Radyushkin:1996nd, Ji:1996nm}. Access to GPDs is typically gained through exclusive reactions, notably deeply virtual Compton scattering~\cite{Ji:1996nm} and deeply virtual meson production~\cite{Radyushkin:1996ru, Collins:1996fb}.

Presently, a series of experiments are being conducted or planned, which can enrich our understanding of the pion internal structure. Prominent among these are the JLab 12 GeV program~\cite{Dudek:2012vr}, the Apparatus for Meson and Baryon Experimental Research (AMBER) at CERN SPS~\cite{Adams:2018pwt}, the forthcoming Electron-Ion Collider (EIC) at Brookhaven National Laboratory~\cite{AbdulKhalek:2021gbh}, and the Electron-Ion Collider in China (EicC)~\cite{Anderle:2021wcy}. These initiatives are designed to probe the pion structure in various kinematic regimes. However, extracting GPDs from experimental data is fraught with challenges, including the chiral-odd nature of certain distributions, such as the transversity GPDs, and the complexity involved in pion production. On the other hand, lattice QCD results, offering complementary insights and potentially guiding experimental efforts, are highly sought after. The advent of Large Momentum Effective Theory (LaMET)~\cite{Ji:2013dva, Ji:2014gla, Ji:2020ect} in 2013 made it possible to directly calculate the Bjorken-$x$ dependence of the parton distributions, extending the scope of GPD studies beyond just the first few Mellin moments~\cite{Hagler:2003jd, Gockeler:2003jfa, Gockeler:2005cj, QCDSF-UKQCD:2007gdl, LHPC:2007blg, Alexandrou:2011nr, Alexandrou:2013joa, Bali:2018zgl, Alexandrou:2019ali, Alexandrou:2022dtc}. This spurred a flurry of lattice research on nucleon and meson GPDs~\cite{Chen:2019lcm, Alexandrou:2020zbe, Lin:2020rxa, Lin:2021brq, Bhattacharya:2022aob, Bhattacharya:2023ays, Lin:2023gxz, Bhattacharya:2023jsc, Holligan:2023jqh} based on the quasi-distribution (see reviews in refs.~\cite{Ji:2020ect, Constantinou:2020hdm}), which is defined in terms of matrix elements of non-local operators involving quark and antiquark fields separated by a spatial distance, rather than a light-cone distance, and thus is calculable on the lattice.

While there are several lattice QCD studies of the nucleon GPDs~\cite{Alexandrou:2020zbe, Lin:2020rxa, Lin:2021brq, Bhattacharya:2022aob, Bhattacharya:2023ays, Bhattacharya:2023jsc, Holligan:2023jqh}, including a few based on alternate methods~\cite{CSSMQCDSFUKQCD:2021lkf,Hannaford-Gunn:2024aix, Dutrieux:2024umu, Bhattacharya:2024qpp}, there have been limited LaMET calculations of the pion GPDs~\cite{Chen:2019lcm, Lin:2023gxz}. Traditionally, GPD calculations are conducted in the Breit (symmetric) frame, requiring the momentum transfer to be symmetrically distributed between the initial and final states. However, this approach incurs substantial computational costs. Recently, a frame-independent method was proposed to extract GPDs from lattice calculations in an arbitrary frame~\cite{Bhattacharya:2022aob}. This innovative approach holds the potential to significantly reduce computational expenses by working in a non-Breit (asymmetric) frame where the initial or final state momentum is fixed.

Two crucial aspects of obtaining the light-cone parton distributions from the lattice calculations are the renormalization and matching process. The GPDs and quasi-GPDs (qGPDs) are typically defined in different renormalization schemes. For the qGPDs extracted from the lattice analyses, renormalization is required to eliminate the ultraviolet (UV) divergences stemming from the Wilson line. Commonly employed renormalization methods include the regularization-independent momentum-subtraction (RI/MOM)~\cite{Constantinou:2017sej, Alexandrou:2017huk, Stewart:2017tvs, Chen:2017mzz} and various ratio schemes~\cite{Radyushkin:2017cyf, Orginos:2017kos, Braun:2018brg, Li:2020xml, Fan:2020nzz}. However, these methods adhere to a factorization relation with the light-cone correlation only at short distances. At long distances, they introduce nonperturbative effects~\cite{Ji:2020brr}, which will impact the qGPDs through the Fourier transform of the matrix elements, thereby affecting the LaMET matching results in Bjorken-$x$ space. To overcome this issue, in this work, we use a hybrid-scheme renormalization~\cite{Ji:2020brr}. The key point of this scheme is to renormalize the matrix elements at short and long distances separately, which can remove the linear divergence at long distances without introducing additional nonperturbative effects.

Furthermore, we match the qGPDs in the lattice renormalization scheme to the light-cone GPDs in the $\overline{\rm MS}$ scheme through LaMET~\cite{Ji:2015qla, Liu:2019urm, Ma:2022ggj, Yao:2022vtp}. During this process, the accuracy of the perturbative calculation plays a significant role in the precision of the final GPD results. In the zero-skewness limit, the matching kernel of GPDs is the same as the one for PDFs~\cite{Liu:2019urm}, which has been derived up to the next-to-next-to-leading order (NNLO)~\cite{Li:2020xml, Chen:2020ody}. We also utilize renormalization group resummation (RGR)~\cite{Gao:2021hxl, Su:2022fiu} to resum the small-$x$ logarithmic terms. 
Additionally, we consider leading-renormalon resummation (LRR)~\cite{Holligan:2023rex, Zhang:2023bxs}, which can remove the renormalon ambiguity in the Wilson-line mass matching, to eliminate the linear power corrections~\cite{Zhang:2023bxs}. Therefore, we achieve a state-of-the-art calculation of the valence pion GPDs using an adapted hybrid-scheme renormalization, along with the implementation of the combined NNLO+RGR+LRR matching.

In this study, we present our lattice calculations of the valence pion GPDs using the LaMET approach, featuring a very small lattice spacing of 0.04 fm within the non-Breit frame. The organization of this paper is as follows: In section~\ref{sec: frame}, we review the definitions of pion GPDs and qGPDs in an arbitrary frame, outline our lattice setup, and describe the methodologies employed to extract the matrix elements from the combined analyses of the two-point and three-point correlation functions. In this section, we also detail the renormalization procedure of the qGPD matrix elements in the coordinate space using the hybrid scheme and discuss the challenges of performing the Fourier transform to get the qGPDs. Subsequently, we apply the LaMET matching approach to derive the valence light-cone GPDs in section~\ref{sec: LC GPDs}. 
This section contains our main results, including the sensitivity of
our results to the perturbative accuracy of the matching and the dependence
on the renormalization scales. Finally, section~\ref{sec: conclusion} provides a summary of our findings.


\section{\label{sec: frame} Valence pion quasi-GPDs in asymmetric frame on the lattice}
\subsection{General considerations}
Our goal is to calculate the valence pion GPDs in an asymmetric frame
using the frame-independent approach laid out in ref.~\cite{Bhattacharya:2022aob}. 
For this, we consider the following non-local iso-vector matrix element 
\begin{equation}
    M^\mu(z, \bar{P}, \Delta) = \left\langle \pi(P^f)|O^{\gamma_\mu}(z)|\pi(P^i) \right\rangle,
\end{equation}
where  $\mu=(t, x, y, z)$ represents the Lorentz indices, $P^i$ and $P^f$ are the initial and final momenta of the pion, respectively, 
$\bar P=(P^i+P^f)/2$, $\Delta=P^f-P^i$, and
\begin{equation}
O^{\gamma_\mu}(z)=\frac{1}{2} \left[\bar{u}(-\frac{z}{2}) \gamma^\mu {\cal W}_{-\frac{z}{2}, \frac{z}{2}} u(\frac{z}{2})-\bar{d}(-\frac{z}{2}) \gamma^\mu {\cal W}_{-\frac{z}{2}, \frac{z}{2}} d(\frac{z}{2}) \right],
\label{def_O}
\end{equation}
where ${\cal W}_{-\frac{z}{2}, \frac{z}{2}}$ is the Wilson line that connects the quark and antiquark fields, ensuring the gauge invariant of the matrix element. 
Here, we use the following normalization of the pion states $\left\langle \pi(P)|\pi(P') \right\rangle=E_p (2\pi)^3 \delta^3(\mathbf{P}-\mathbf{P'})$.
Following ref.~\cite{Bhattacharya:2022aob}, we write the matrix element in terms of the Lorentz-invariant amplitudes $A_i(z\cdot\bar{P}, z\cdot\Delta, \Delta^2, z^2)$ for $i=1,2,3$, as follows:
\begin{equation}
	M^\mu(z,\bar P, \Delta) = \bar P^\mu A_1 + m_{\pi}^2 z^\mu A_2 + \Delta^\mu A_3.
	\label{eq: param}
\end{equation}

The coordinate space light-cone valence pion GPDs $H$ is usually defined in terms of $M^+(z,\bar{P},\Delta)$ as
$M^+(z,\bar P,\Delta)=P^+ H(z,\bar P,\Delta)$, i.e. the momentum space light-cone GPDs is written
as
\begin{equation}
    H(x,\xi,t)=P^+ \int \frac{d z^-}{2 \pi} e^{i z^- P^+ x} H(z,\bar P, \Delta),
    \label{GPD_def}
\end{equation}
where the skewness parameter $\xi=-\Delta^+/(2P^+)$ and $t=\Delta^2$. 
However, it is possible to define the light-cone GPDs in a frame-independent, i.e., Lorentz-invariant way~\cite{Bhattacharya:2022aob} as
\begin{equation}
H_{\rm LI}(z\cdot\bar{P}, z\cdot\Delta, \Delta^2, 0) = A_1(z\cdot\bar{P}, z\cdot\Delta, \Delta^2, 0) + \frac{z \cdot \Delta}{z \cdot \bar{P}} A_3(z\cdot\bar{P}, z\cdot\Delta, \Delta^2, 0).
\end{equation}
Motivated by this, the Lorentz-invariant definition of qGPDs can be written as
\begin{equation}
	\widetilde{H}_{\rm LI}(z\cdot\bar{P}, z\cdot\Delta, \Delta^2, z^2) \equiv A_1(z\cdot\bar{P}, z\cdot\Delta, \Delta^2, z^2) + \frac{z \cdot \Delta}{z \cdot \bar{P}} A_3(z\cdot\bar{P}, z\cdot\Delta, \Delta^2, z^2).
\end{equation}
This is a natural choice for the qGPDs because in the light-cone limit $z^2\rightarrow0$, it should be equal to the light-cone GPDs up to the leading order in $\alpha_s$
\begin{equation}
    \lim_{z^2\rightarrow 0} \widetilde{H}_{\rm LI}(z\cdot\bar{P}, z\cdot\Delta, \Delta^2, z^2) = H_{\rm LI}(z\cdot\bar{P}, z\cdot\Delta, \Delta^2, 0) + {\cal O}(\alpha_s).
\end{equation}
In the Lorentz-invariant framework, the momentum-space GPD is defined as~\cite{Bhattacharya:2022aob}
\begin{equation}
    H(x,\xi,t)=\int \frac{d (z\cdot \bar{P})}{2 \pi} e^{i x z\cdot \bar{P}} H_{\rm LI}(z\cdot \bar{P},-2 \xi (z\cdot \bar{P}), t,0),
    \label{GPD_LI}
\end{equation}
and similarly for qGPD
\begin{equation}
    \widetilde H(x,\xi,t,\bar{P}_z)=\int \frac{d (z \cdot \bar{P})}{2 \pi} e^{ix z \cdot \bar{P}} \widetilde H_{\rm LI}(z\cdot \bar{P},-2 \xi (z\cdot \bar{P}), t,z^2).
    \label{qGPD_LI}
\end{equation}

By studying the behavior under hermiticity and time-reversal transformations simultaneously, it was found that the amplitude $A_3$ is an odd function of $z\cdot \Delta=-2\xi(z\cdot\bar{P})$~\cite{Bhattacharya:2022aob, Bhattacharya:2023jsc}.
In the forward limit, GPDs should smoothly approach PDFs, implying that $A_3$ should be equal to 0 in this work at zero skewness. 
In the following subsections, we will discuss our lattice setup and lattice QCD results on $A_i$.

\subsection{Lattice setup}
Our lattice calculations use the gauge ensembles provided by the HotQCD collaboration~\cite{Bazavov:2019www}, utilizing a 2+1 flavor setup with Highly Improved Staggered Quark (HISQ) action~\cite{Follana:2006rc}. The lattice configuration has dimensions of $N_s\times N_t=64^3 \times 64$ and a lattice spacing of $a=0.04$ fm. The sea quark masses are adjusted to yield a pion mass of 160 MeV. In the valence sector, we use the Wilson-Clover action with 
one level of hypercubic (HYP) smearing~\cite{Hasenfratz:2001hp}. For the coefficient of the clover term, we utilize the tree-level tadpole-improved value computed with the fourth root of the plaquette, yielding 1.02868~\cite{Gao:2020ito}. The valence quark masses in the Wilson-Clover action are tuned to $am_q=-0.033$, resulting in a pion mass of 300 MeV. While mixed action and partially quenched setups can theoretically introduce larger lattice artifacts compared to unitary actions, previous studies (e.g., refs.~\cite{Bhattacharya:2015wna, Mondal:2020cmt, Gao:2020ito, Gao:2022iex}) have shown that these effects are milder than other systematic or statistical uncertainties in practice.

\begin{table}
\rowheight{1.5}
    \resizebox{\textwidth}{!}{
    \begin{tabular}{c|cccccccc}
    \hline
    Frame & $t_s/a$ & $\mathbf{n}^f=(n^f_x, n^f_y, n^f_z)$ & $m_z$ & $P_z$[GeV] & $\mathbf{n}^{\Delta}=(|n^\Delta_x|, |n^\Delta_y|, |n^\Delta_z|)$ & $-t$[GeV$^2$] & \#cfgs & (\#ex, \#sl) \\
    \hline
    Breit & 9,12,15,18 & (1, 0, 2) & 2 & 0.968 & (2, 0, 0) & 0.938 & 115 & (1, 32) \\
    \hline
    \multirow{4}{*}{non-Breit} & 9,12,15,18 & (0,0,0) & 0 & 0 & (0,0,0) & 0 & 314 & (3, 96) \\
    \cline{2-9}
    & 9,12,15,18 & (0,0,2) & 2 & 0.968 & (1,2,0) & 0.952 & 314 & (4, 128) \\
    \cline{2-9}
    & 9,12,15 & (0,0,3) & 2 & 1.453 & \multirow{2}{*}{\thead{\big[(0,0,0), (1,0,0) \\ $\,\;$(1,1,0), (2,0,0)\\ $\quad$(2,1,0), (2,2,0)\big]}} & \thead{[0, 0.229, 0.446,$\quad$ \\ 0.855, 1.048, 1.589]} & 314 & (4, 128) \\
    \cline{2-5}\cline{7-9}
    & 9,12,15 & (0,0,4) & 3 & 1.937 & & \thead{[0, 0.231, 0.455,$\quad$ \\ 0.887, 1.095, 1.690]} & 564 & (4, 128) \\
    \hline
    \end{tabular}%
    }
    \vspace{5pt}
	\caption{\label{tb: statistics} Details of the measurements and statistics for the Breit and non-Breit frames are shown. The symbol $t_s$ represents the source-sink separation. We present the momentum in units of $2\pi/(aN_s)$,  including the final momentum ($\mathbf{n}^f$) and the boost momentum along the $z$-direction ($m_z$). $P_z$ denotes the physical values of the momentum in the $z$-direction. For the momentum transfer, we show both the three-dimensional lattice unit momentum transfer ($\mathbf{n}^{\Delta}=\mathbf{n}^f-\mathbf{n}^i$), where $\mathbf{n}^i$ denotes the initial momentum, and the 4-dimensional physical unit momentum transfer ($t\equiv\Delta^2$). Additionally, we provide the numbers of gauge configurations (\#cfgs) as well as the counts for both exact (\#ex) and sloppy (\#sl) inversion samples per configuration.} 
\end{table}

Central to our computational approach is the use of momentum-boosted smeared sources~\cite{Bali:2016lva}. The quark propagators are obtained through the application of the multigrid algorithm~\cite{Brannick:2007ue} to invert the Wilson-Dirac operator using the QUDA software suite
~\cite{Clark:2009wm, Babich:2011np, Clark:2016rdz} on GPUs. 
In this work, we consider the pion boosted along the $z$-direction 
with momenta
$P_z = 2\pi n_z/(aN_s)$ with $n_z$ being an integer. 
To obtain a good signal for the boosted pions,
momentum-boosted sources and sinks are constructed using a Gaussian profile, with boost momenta $k_z=2\pi m_z/(aN_s)$ in the $z$-direction. Source construction is done in the Coulomb gauge, and the Gaussian profile is created with the radius of 0.208 fm~\cite{Izubuchi:2019lyk, Gao:2020ito}. Employing these quark propagators, we have computed both the two-point and three-point hadron correlation functions. To increase statistics per configuration, we combined multiple exact (high-precision) and sloppy (low-precision) samples and implemented the all-mode averaging (AMA) technique~\cite{Shintani:2014vja}. The lattice parameters used in this study are summarized in table~\ref{tb: statistics}. 

We can get the matrix elements by analyzing the two-point and three-point correlation functions. The two-point correlation function is defined as
\begin{equation}
	C_{{\rm 2pt}}(\mathbf{P}, t_s) = \sum_{\mathbf{x}} e^{-i\mathbf{P}\cdot\mathbf{x}} \left\langle \pi_{\rm s}(\mathbf{x}, t_s)\pi_{\rm s}^\dagger(\mathbf{0}, 0) \right\rangle,
	\label{eq: c2pt}
\end{equation}
where $\mathbf{P}$ denotes the spatial momentum, $\mathbf{x}$ and $\mathbf{0}$ represent the spatial coordinates, and $t_s$ and $0$ correspond to the time coordinates. Here, $\pi^\dagger_s$ and $\pi_s$
stand for the pion creation and annihilation operators, respectively, with the subscript ``${\rm s}$" indicating ``smeared".
The three-point correlation function is defined as
\begin{equation}
	C_{{\rm 3pt}}^{\mu}(\mathbf{P}^f, \mathbf{q}; t_s, \tau; z) = \sum_{\mathbf{x,y}} e^{-i\mathbf{P}^f\cdot\mathbf{x}} e^{i\mathbf{q}\cdot\mathbf{y}} \left\langle \pi_{\rm s}(\mathbf{x}, t_s)O^{\gamma_\mu}(\mathbf{y}, \tau; z)\pi_{\rm s}^\dagger(\mathbf{0}, 0) \right\rangle,
	\label{eq: c3pt}
\end{equation}
where $\mathbf{P}^f=2 \pi\mathbf{n}^f/(aN_s) $ and $\mathbf{q}$ denote the spatial final momentum and momentum transfer, respectively. The spatial initial momentum is given by $\mathbf{P}^i=\mathbf{P}^f-\mathbf{q}$. The quark bilinear operator, $O^{\gamma_\mu}(z)$, defined in eq.~(\ref{def_O}), is also characterized by its spacetime insertion position ($\mathbf{y}, \tau$). In this work, we only consider the case of zero skewness, that is, $n_z^{\Delta}=0$ and $P_z=P_z^f=P_z^i$.
For each value of $P_z$, we consider several values of 
$\mathbf{q}=2\pi\mathbf{n}^{\Delta}/(aN_s)$. The different choices of momenta used in our study are summarized in table~\ref{tb: statistics}, including the momentum transfer $t$ in physical units.
As one can see from the table, for the smallest non-zero value of $P_z$, we have performed calculations in both Breit and non-Breit frames at very similar values of $t$. 

The approach used to obtain the pion matrix element here closely follows our previous works~\cite{Gao:2020ito, Gao:2021dbh, Gao:2021xsm, Gao:2022iex, Gao:2022vyh, Ding:2024lfj}. We fit the large $t_s$ behavior of the two-point correlation functions with two or three states and obtain the energies of these states for different momenta. The corresponding energy levels are then used in the analyses of the three-point function and the extraction
of the pion matrix elements $M^{\mu}$. We consider the following ratio of 
\begin{align}
    R^{\mu}(\mathbf{P}^f, \mathbf{P}^i; t_s, \tau;z) &\equiv \sqrt{E_0^f E_0^i}\frac{C_{{\rm 3pt}}^{\mu}(\mathbf{P}^f, \mathbf{P^i}; t_s, \tau;z)}{C_{{\rm 2pt}}(\mathbf{P}^f, t_s)}\nonumber\\
    &\qquad \times \left[ \frac{C_{{\rm 2pt}}(\mathbf{P}^i, t_s - \tau)C_{{\rm 2pt}}(\mathbf{P}^f, \tau)C_{{\rm 2pt}}(\mathbf{P}^f, t_s)}{C_{{\rm 2pt}}(\mathbf{P}^f, t_s - \tau)C_{{\rm 2pt}}(\mathbf{P}^i, \tau)C_{{\rm 2pt}}(\mathbf{P}^i, t_s)} \right]^{1/2},
	\label{eq: ratio}
\end{align}
where $E_0^i$ and $E_0^f$ correspond to the ground-state pion energies in the initial and final states, respectively, and the kinematic factor $\sqrt{E_0^f E_0^i}$ is used for the normalization. For large $t_s$ and $\tau$, this ratio gives the matrix element $M^{\mu}$. We assume that for large $t_s$ and $\tau$, two or three states mainly contribute to $R^{\mu}(\mathbf{P}^f, \mathbf{P}^i; t_s, \tau;z)$ and perform the corresponding fits to obtain the matrix element. More details of the fits are provided in appendix~\ref{sec: fit results}.

\subsection{Lattice results on the Lorentz-invariant amplitudes}

Having determined the matrix elements $M^{\mu}$ for different values of the kinematic variables in eq.~(\ref{eq: param}), we can deduce the Lorentz-invariant amplitudes $A_i$. The results of $A_i$ obtained from the Breit and non-Breit frames with comparable values of $-t$ should be consistent. Therefore, we select two sets of lattice data from different frames, as shown in the first and third rows of table~\ref{tb: statistics}, which have identical values of $P_z$ and very close $-t$ values. The dataset from the Breit frame has $-t=0.938$ GeV$^2$, while the dataset from the non-Breit frame has $-t=0.952$ GeV$^2$. According to eq.~(\ref{eq: param}) and $z^\mu=(0,0,0,z)$ in our case, for the non-Breit frame, we can simultaneously solve $A_1$ and $A_3$ by combining the data of $\gamma_t$ and $\gamma_\perp$ components. For the Breit frame with $\Delta^t=0$ and $\bar{P}_\perp=0$, we can directly get $A_1$ and $A_3$ from $\gamma_t$ and $\gamma_\perp$ components, respectively. 

\begin{figure}[t]
	\centering
	\includegraphics[width=0.32\linewidth]{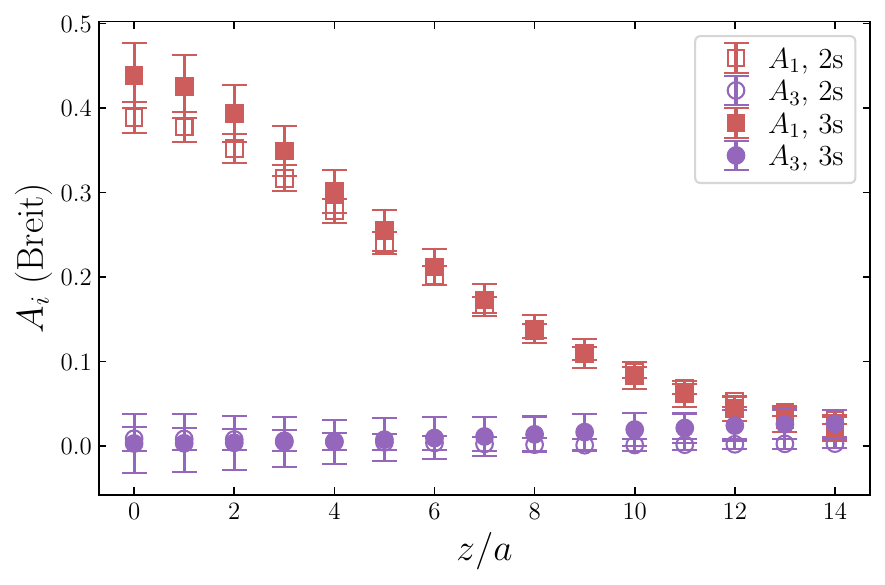} 
	\includegraphics[width=0.32\linewidth]{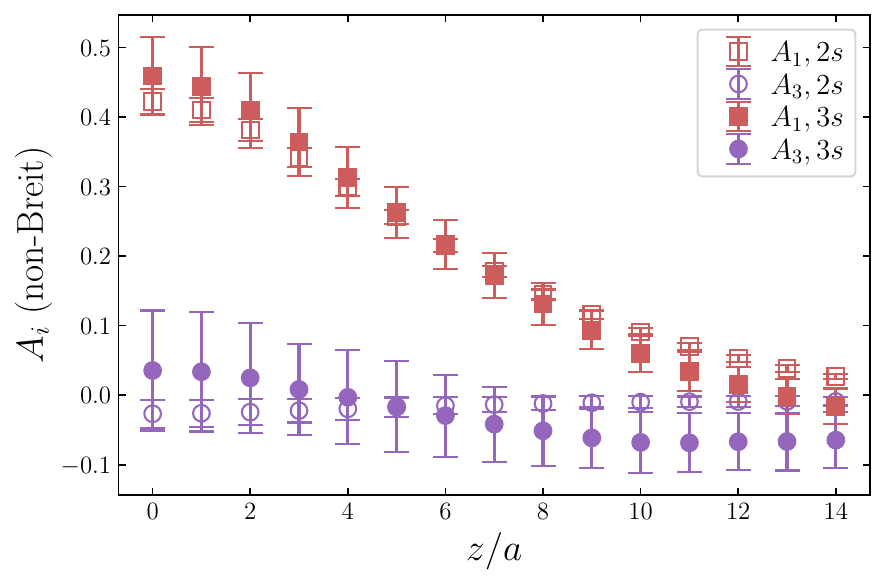} 
	\includegraphics[width=0.32\linewidth]{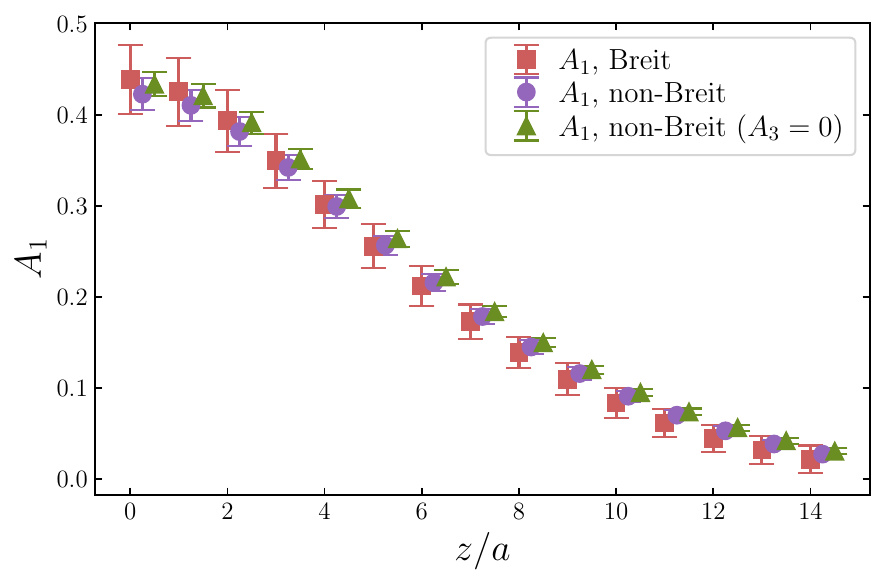} 
	\caption{\label{fig: amplitude} The results for the amplitudes $A_i(i=1,3)$, obtained from the Breit frame (left panel) and non-Breit frame (middle panel) with identical $P_z=0.968$ GeV and very similar transfer momenta, are shown, including the outcomes of both the 2-state and 3-state fits. In the right panel, we compare the results of $A_1$ obtained from the Breit frame using the 3-state fit with two different determinations of $A_1$ obtained from the non-Breit frame using the 2-state fits with and without setting $A_3$
    to zero, see text.}
\end{figure}

The amplitudes obtained from these two datasets with $P_z=0.968$ GeV are shown in figure~\ref{fig: amplitude}. The left two panels show the results obtained using both 2-state and 3-state fits within the Breit and non-Breit frames, respectively. The 2-state fit results are represented by the open symbols, while the 3-state fit results are denoted as the filled symbols, labeled as ``2s" and ``3s" in the figure, respectively. One can see that the contamination from the excited states is apparent in the Breit frame, so the 3-state fit is needed. For the non-Breit frame, it can be seen that these two kinds of fit results are comparable within 1-$\sigma$ error, and the 2-state fit results are more stable. In both frames, we find that $A_3$ is zero within error, as expected. In the right panel of figure~\ref{fig: amplitude}, we investigate the frame independence of $A_1$ by selecting the 3-state fit results for the Breit frame and the 2-state fit results for the non-Breit frame. Furthermore, since we observe that $A_3$ is compatible with zero within errors also for the non-Breit frame, we perform the extraction of $A_1$ from the non-Breit frame based on setting $A_3=0$ and using the $\gamma_t$ component. The results from this analyses are shown as the green triangles in the right panel of figure~\ref{fig: amplitude}. As depicted in the figure, the non-Breit frame results, both with and without setting $A_3=0$, exhibit good agreement and also are consistent with the Breit frame results.

Therefore, we conclude that it is possible to obtain $A_1$ from the calculations in the non-Breit frame by initially setting $A_3$ to zero.
Specifically, since in our lattice QCD calculations, both the initial and final pion states are boosted along the $z$ direction by an equal amount $P_z$, the Lorentz-invariant definition for the coordinate-space qGPDs for $\xi=0$ can be directly expressed as
\begin{equation}
    \widetilde{H}_{\rm LI}(zP_z, \Delta^2, z^2) = A_1 (zP_z, \Delta^2, z^2) = M^t(zP_z, \Delta^2, z^2)/\bar{P}^t.
\label{eq: LI qGPD}
\end{equation}
For convenience, we denote the dimensionless matrix element as $M\equiv M/\bar{P}$ hereafter.

Since we aim to get the valence pion GPDs within the LaMET framework, we selected the two largest available momenta in the non-Breit frame for our calculations, as detailed in the last two rows of table~\ref{tb: statistics}. Additionally, these two datasets can be used to study the $P_z$ dependence of the final light-cone GPDs, which can reflect the effectiveness of our perturbative matching, as discussed later. For each momentum, we analyzed six values of the momentum transfer $-t$. Using the matrix elements we have obtained, several steps are required to derive the qGPDs, including renormalization, extrapolation for the large region of $z$, and the Fourier transform. Subsequently, the light-cone GPDs can be obtained from the qGPDs using the LaMET.

\subsection{\label{sec: renormalization}Renormalization}
As we mentioned before, the renormalization of the qGPD matrix elements is crucial for removing the UV divergences from the Wilson line as well as for matching to
the light-cone GPDs, which are usually defined in the $\overline{\rm MS}$ scheme. Since the non-local quark bilinear operator is multiplicatively renormalizable, if we call the matrix elements extracted directly from the lattice calculations as the bare matrix elements, the relation between the bare and renormalized matrix elements can be expressed as~\cite{Ji:2017oey, Ishikawa:2017faj, Green:2017xeu, Ji:2020brr}
\begin{equation}
	M^{\rm B}(z, a) = Z(a) e^{-\delta m(a)|z|} e^{-\bar{m}_0|z|} M^{\rm R}(z),
\label{eq: renorm}
\end{equation}
where the superscripts ``B" and ``R" denote the ``bare" and ``renormalized" quantities, respectively, $Z(a)$ includes the $z$-independent logarithmic divergence, the term with $\delta m(a)$ accounts for the linear divergence, and the term with $\bar{m}_0$ is introduced to address the scheme dependence of $\delta m$ and to match the lattice scheme to the $\overline{\rm MS}$ scheme~\cite{LatticePartonCollaborationLPC:2021xdx, Gao:2021dbh, Holligan:2023rex, Zhang:2023bxs}. Theoretically, $\bar{m}_0$ should be a constant and independent of $z$.

The hybrid scheme renormalization is defined by merging the ratio scheme for short distances with the explicit subtraction of self-energy divergences in the Wilson line for long distances~\cite{Ji:2020brr}
\begin{equation}
    M^{\rm R}(z, z_s; P_z, t) =
    \left\lbrace 
    \begin{aligned}
		&\frac{M^{\rm B}(z, P_z, t)}{M^{\rm B}(z, 0, 0)}, &|z| \leq |z_s|; \\
		&\frac{M^{\rm B}(z, P_z, t)}{M^{\rm B}(z_s, 0, 0)} e^{(\delta m+\bar{m}_0)|z-z_s|}, &|z| > |z_s|.
    \end{aligned}\right.
\end{equation}
Here, $z_s$ represents the position where the ratio scheme is matched onto the explicit subtraction of the divergence in the Wilson line, which is part of the renormalization scheme. To reduce artifacts and improve the signal, we can further divide $M^{\rm R}$ by the renormalized electromagnetic form factor $F(P_z, t) \equiv M^{\rm B}(0, P_z, t)/M^{\rm B}(0, 0, 0)$ such as
\begin{equation}
    M^{\bar{\rm R}}(z, z_s; P_z, t) = M^{\rm R}(z, z_s; P_z, t)/F(P_z, t).
\end{equation}
Notably, we need to multiply $F(P_z, t)$ back into our results when calculating the pion qGPDs, as presented in the next subsection.

At short distances, all the divergences can be directly canceled by such a ratio. However, for long distances, in order to perform the renormalization, we first need to determine the values of $\delta m$ and $\bar{m}_0$. In line with our previous studies, we determine $\delta m$ using lattice QCD results on the static quark-antiquark potential and the free energy of a static quark at non-zero temperatures~\cite{Bazavov:2018wmo}: $a\delta m=0.1508(12)$ for $a=0.04$ fm lattice~\cite{Gao:2021dbh}. The value of $\bar{m}_0$ can be obtained using the bare matrix element results at zero momentum and zero momentum transfer. Namely, by comparing the lattice computations at $P_z=0$ GeV and $t=0$ GeV$^2$ with their corresponding $\overline{\rm MS}$ Operator Product Expansion (OPE) expressions for such a ratio
\begin{align}\label{eq: m0}
	e^{(\delta m+\bar{m}_0)\delta z} \frac{M^{\rm B}(z+\delta z)}{M^{\rm B}(z)} &=  \frac{C_{0}(\alpha_{s}(\mu_0(z+\delta z)), \mu_0^2(z+\delta z)^2)}{C_{0}(\alpha_{s}(\mu_0(z)), \mu_0^2z^2)}\nonumber\\
    &\qquad \times\exp\left[\int_{\alpha_s(\mu_0(z+\delta z))}^{\alpha_s(\mu_0(z))} {d\alpha_s( \mu')\over \beta[\alpha_s(\mu')]}\gamma_O[\alpha_s(\mu')]\right],
\end{align}
where $\gamma_O$ is the anomalous dimension of the quark bilinear operator, known up to three loops~\cite{Braun:2020ymy}, and the expressions for $\alpha_s$ and $\beta$ can be found in appendix~\ref{sec: matching}, we can extract the values of $\bar{m}_0$. Here, we set $|\delta z|/a=1$. The use of the zero momentum matrix element results in only the 0-th Wilson coefficient $C_0$ remaining, which is calculated up to the NNLO~\cite{Li:2020xml, Chen:2020ody}. Furthermore, we have incorporated RGR~\cite{Gao:2021hxl} at the next-to-next-to-leading-logarithm (NNLL) accuracy. The RGR allows us to effectively sum the large logarithmic terms arising in the perturbative expansions and to evolve the running coupling from the physical scale $\mu_0 = 2\kappa e^{-\gamma_E}/|z|$ to the factorization scale $\mu$ = 2 GeV. Here, $\kappa$ is a proportionality constant, and $\gamma_E$ is the usual Euler constant. In principle, we should choose $\kappa\sim 1$ to match the natural physical scale $1/|z|$, but we found that eq.~(\ref{eq: m0}) with $\kappa < 1$ cannot describe the lattice data with a constant $\bar{m}_0$ in $z$. Therefore, to estimate the scale variation uncertainty, we choose our central value of $\kappa$ as 1.414, which is still close to 1, and vary it between 1 and 2, where the strong coupling constant remains reasonably small ($\alpha_{s}<$ 0.3) at short distances we considered. More details about the process for selecting the range of $\kappa$ can be found in appendix~\ref{sec: kappa}.
Additionally, we include LRR~\cite{Zhang:2023bxs, Holligan:2023rex} in the Wilson coefficient that captures the dominant contributions from renormalons and improves the convergence properties of the perturbative series~\cite{Zhang:2023bxs}.

\begin{figure}[t!]
	\centering
    \vspace{-20pt}
    \includegraphics[width=0.7\textwidth]{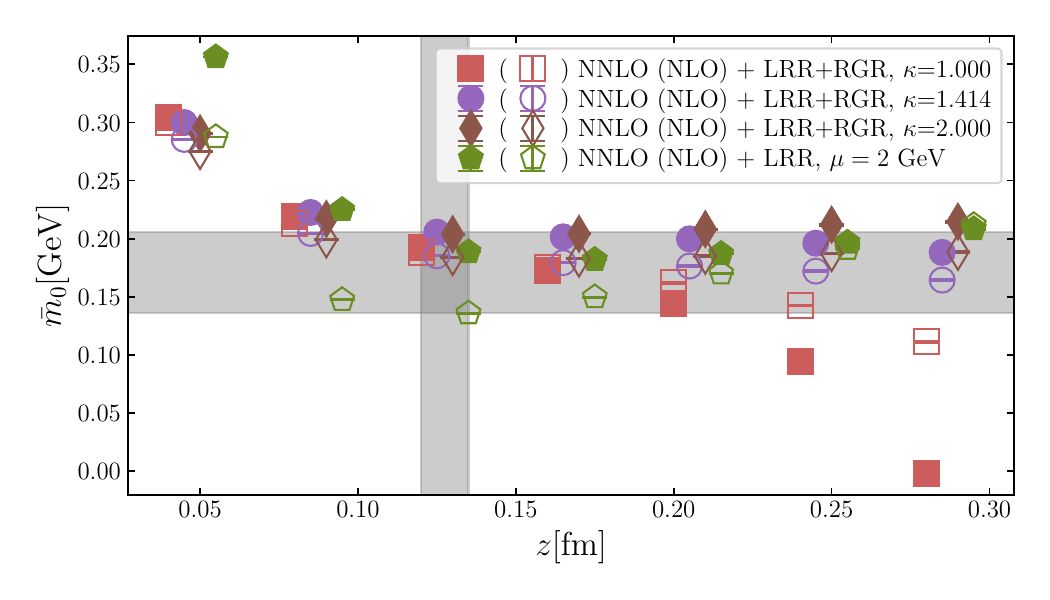}
	\caption{ The additional Wilson line renormalization mass parameter $\bar{m}_0$, obtained using NNLO(NLO)+LRR and NNLO(NLO)+LRR+RGR  expressions for $C_0$ at different values of $z$. The filled symbols correspond to NNLO results for $C_0$, while the open symbols correspond to NLO results for $C_0$. The horizontal band indicates the range of values of $\bar m_0$ used in our analysis. The vertical indicates the value of $z$ at which $\bar m_0$ was determined.}
\label{fig: m0}
\end{figure}

In figure~\ref{fig: m0}, we show the results of $\bar{m}_0$ obtained with Wilson coefficients at the accuracies of NNLO(NLO)+LRR and NNLO(NLO)+LRR+RGR. To enhance visual clarity, results with varying scale settings of the same order are plotted with a slight horizontal offset. As mentioned before, $\bar{m}_0$ is excepted to be a constant and independent of $z$. However, in practice, this is not always the case for all $z$ values, as illustrated in figure~\ref{fig: m0}. For the two smallest values of $z$, we observe lattice artifacts in the determination of $\bar{m}_0$, as expected. In contrast, at $\kappa =1.414$ and $2$ the dependence of $\bar{m}_0$ on $z$ becomes much milder for larger $z$ values until $z\sim 0.3$ fm where perturbation theory breaks down. The only exception occurs when $\kappa=1$, where the running coupling already becomes very large for $z>0.2$ fm, as indicated in figure~\ref{fig: alpha allkappa} in appendix~\ref{sec: kappa}, causing the perturbative expansion of the Wilson coefficient to break down.
The values of $\bar{m}_0$ determined using NLO+LRR also show some dependence on $z$ due to missing higher-order corrections. However, thanks to LRR, the dependence of $\bar{m}_0$ on $z$ is significantly reduced already at this level compared to our prior analyses of the pion PDFs~\cite{Gao:2021dbh}.
All other determinations of $\bar{m}_0$ obtained with $\kappa \ge 1.414$ are consistent with each other and show very small $z$-dependence for $z>0.08$ fm. Since the results at $z=0.12$ fm show better consistency among all the strategies, we select the $\bar{m}_0$ results at $z=0.12$ fm, as indicated by the intersection area of the gray bands in figure~\ref{fig: m0}, to carry out the hybrid-scheme renormalization. For example, $\bar{m}_0=0.1922(3)$ GeV at NNLO+LRR+RGR accuracy with $\kappa=1$. Actually, the slight variation in the choice of $\bar{m}_0$ has minimal impact on the renormalized results, as illustrated in appendix~\ref{sec: z dep of m0}. We also choose $z_s=0.12$ fm as our default choice for the hybrid scheme.

\subsection{\label{sec: FT}Large $z$ extrapolation and Fourier transform}

\begin{figure}[t]
	\centering
    \vspace{-30pt}
	\includegraphics[width=0.7\textwidth]{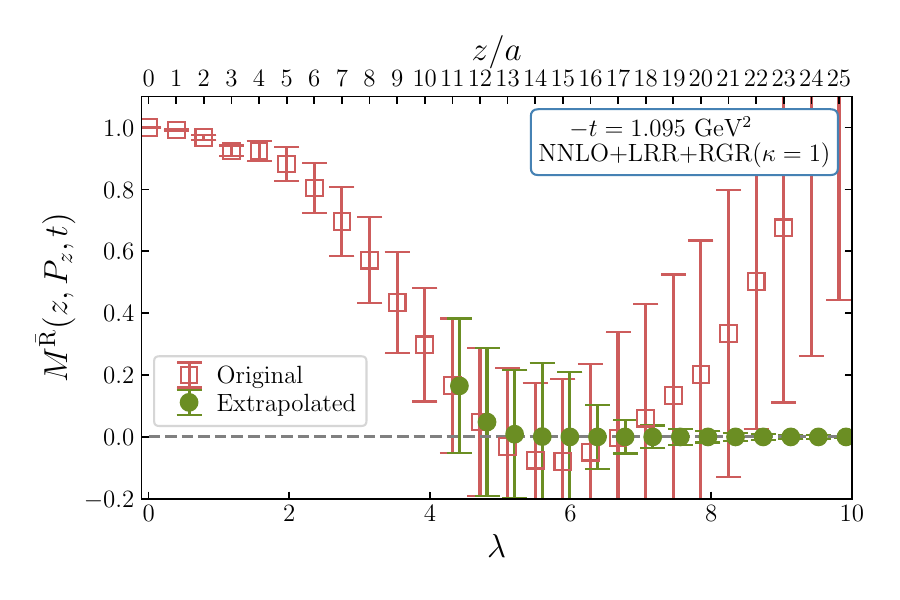}
	\caption{\label{fig: extrapolated} The extrapolated results, along with the original renormalized matrix elements for the case of $-t=1.095$ GeV$^2$, obtained with NNLO+LRR+RGR($\kappa=1$) coefficients, are presented as a function of $\lambda=zP_z$.}
\end{figure}
In this work, our aim is to get the valence light-cone GPDs in momentum ($x$) space.
In the isospin symmetric limit, the valence light-cone GPDs, as well as qGPDs, are equal to the isovector GPDs, as discussed in ref.~\cite{Izubuchi:2019lyk}.
To obtain the momentum-space isovector (valence) qGPDs, we need to perform a Fourier transform on $\lambda=z P_z$
of the coordinate-space isovector qGPDs
\begin{equation}
    \widetilde{H}(x, P_z, t)
    = 2\int_{-\infty}^{\infty} \frac{{\rm d}\lambda}{2\pi} e^{ix\lambda} \widetilde{H}_{\rm LI}(z P_z, t,z^2)
    = F(P_z,t) \int_{-\infty}^{\infty} \frac{{\rm d}\lambda}{\pi} e^{ix\lambda} M^{\bar{\rm R}}(z, P_z, t), 
\end{equation}
where $\widetilde{H}(x, P_z, t)$ is the valence pion qGPDs, and the factor of 2 at the beginning of the middle formula comes from the definition of $O^{\gamma_{\mu}}$ in eq.~(\ref{def_O}) and the fact that the integral of the valence $u$ quark distribution in the momentum space from $0$ to $1$ should be one.

In figure~\ref{fig: extrapolated}, we take the case of $-t=1.095$ GeV$^2$ as an example to present our results of the renormalized matrix elements, denoted by the red square symbols, as a function of $\lambda$. 
Due to the large statistical errors at large $z$ and the finite volume effects, we can reliably calculate the renormalized matrix elements only up to a certain value of $\lambda$. This prevents us from performing the Fourier transform directly on $M^{\bar{\rm R}}$. 
To address this problem, we perform an extrapolation of the renormalized matrix elements from a truncation position $z_t$ and replace the lattice results with the extrapolated values for $z\ge z_t$.
We expect that $M^{\bar{\rm R}}$ should vanish exponentially at large $z$~\cite{Ji:2020brr, Gao:2021dbh}, which means that the extrapolation mainly affect the small-$x$ region of GPDs which is beyond the region of LaMET prediction~\cite{Gao:2021dbh}.
Therefore, we fit the lattice results of $M^{\bar{\rm R}}$ at large $z$ using the following ansatz
\begin{equation}
    M^{\bar{\rm R}} = A\frac{e^{-mz}}{(zP_z)^d},
    \label{eq: decay model}
\end{equation}
where $\{A,m,d\}$ are the fit parameters. In practice, we choose $N$ data points to the left from a critical position $z_c$ where the matrix element results become unreliable, such as showing negativity or lacking a decaying trend, to do the fit. Subsequently, employing the decay model~(\ref{eq: decay model}) with the fitted parameters, we simulate results starting from a truncation position $z_t/a=z_c/a-2$, which is the best choice to ensure the extrapolation start from large enough distance, and extend towards long distances where the matrix element approaches zero. We constrain the fit parameters with the priors $\{A, m\}>0$ and extrapolate the data to $z/a=64$, which can lead to extrapolated results decaying to zero in most cases. However, in the cases of $-t=[0, 0.231]$ GeV$^2$ for $P_z=1.937$ GeV, due to the slower decay, we need to impose a stricter constraint $m>0.2$ GeV~\cite{Gao:2021dbh} as a fit prior and extrapolate the matrix elements to a longer distance $z/a=200$. For a clearer description, we display the extrapolated results by the green circles in figure~\ref{fig: extrapolated}, which are calculated using the fitted parameters obtained with $N$ = 3. A more detailed discussion of the fit range comparison is available in appendix~\ref{sec: extrapolation}. As illustrated, the extrapolated results exhibit expected decay behavior and approach zero as the distances become sufficiently large. Therefore, the extrapolation can effectively correct the artificial behavior of the lattice data at large $\lambda$.

\begin{figure}[t!]
	\centering
    \vspace{-30pt}
	\includegraphics[width=0.48\textwidth]{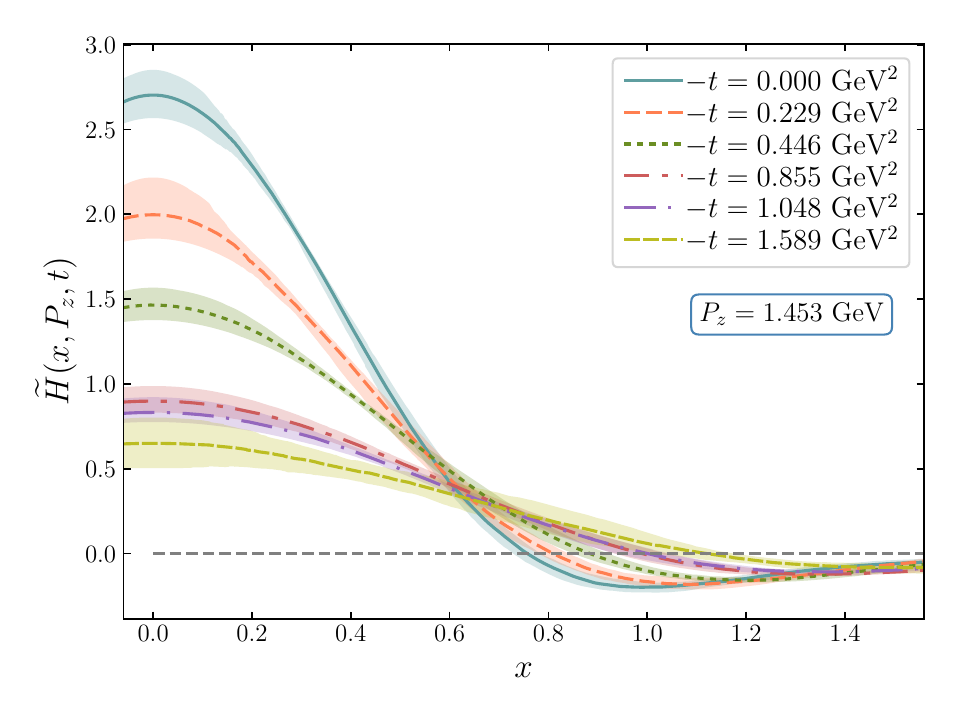}
	\includegraphics[width=0.48\textwidth]{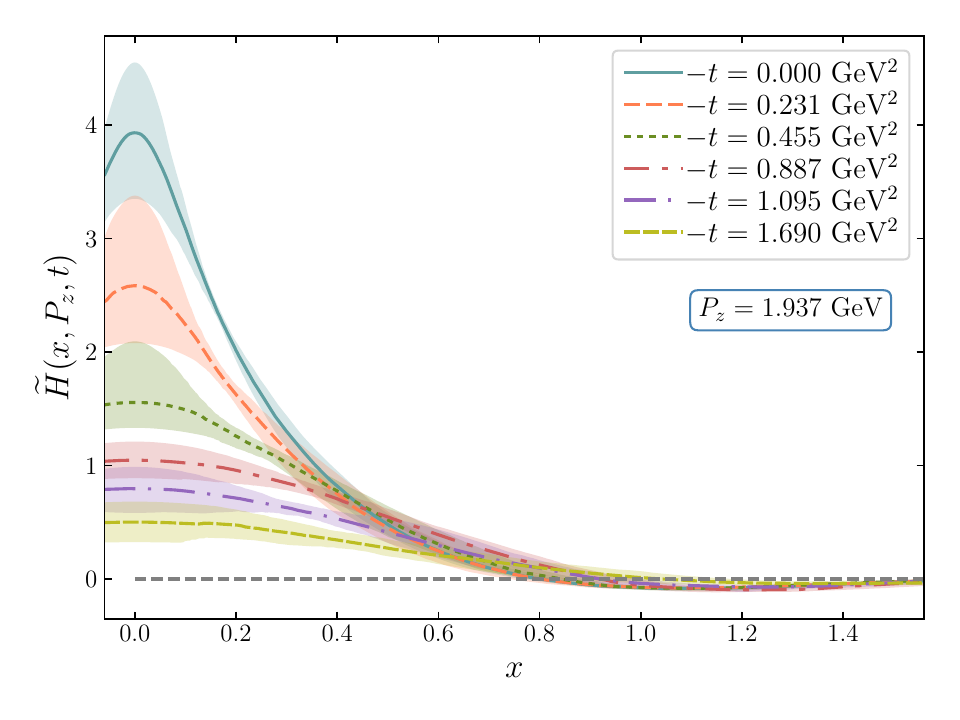}
	\caption{\label{fig: FT} The pion valence quark qGPD results for all $-t$ values obtained with NNLO+LRR+RGR ($\kappa=1.414$) coefficient are shown as a function of the Bjorken-$x$. Left: $P_z=1.453$ GeV, right: $P_z=1.937$ GeV.}
\end{figure}

To obtain the $x$ dependence of the qGPDs, we need to integrate the extrapolated renormalized matrix elements over all values of $\lambda=zP_z$ in the Fourier transform.  
In figure~\ref{fig: FT}, we show the valence pion qGPD results at $P_z=1.453$ and 1.937 GeV, which are obtained using NNLO+LRR+RGR ($\kappa=1.414$) renormalization. We present results for six $-t$ values corresponding to each $P_z$ value. We see that as $-t$ increases, the distributions start at smaller values and decay slower with increasing $x$. If we compare the data with similar $-t$ sizes between these two panels, we can find that there are considerable differences between $P_z=1.453$ GeV and $1.937$ GeV results. We expect that the perturbative matching to light-cone GPDs, discussed in the next section, can correct for a significant portion of these differences. 

\section{\label{sec: LC GPDs} Numerical results on the valence pion light-cone GPDs}
We utilize the LaMET approach for the perturbative matching of qGPDs to the light-cone GPDs
\begin{equation}
	\begin{aligned}
		H(x, \mu, t) &=
        \int_{-\infty}^{\infty} \frac{{\rm d}k}{|k|} \int_{-\infty}^{\infty} \frac{{\rm d}y}{|y|} {\cal C}_{\rm evo}^{-1}\left( \frac{x}{k}, \frac{\mu}{\mu_0} \right) {\cal C}^{-1}\left( \frac{k}{y}, \frac{\mu_0}{yP_z}, |y|\lambda_s \right) \widetilde{H}(y, P_z, t, z_s, \mu_0) \\
        &\approx \sum_{i,j} \frac{\Delta k}{|k_i|}\frac{\Delta y}{|y_j|} {\cal C}_{\rm evo}^{-1}\left( \frac{x}{k_i}, \frac{\mu}{\mu_0} \right) {\cal C}^{-1}\left( \frac{k_i}{y_j}, \frac{\mu_0}{y_jP_z}, |y_j|\lambda_s \right) \widetilde{H}(y_j, P_z, t, z_s, \mu_0),
	\end{aligned}
\label{eq: perturbative matching}
\end{equation}
where $\lambda_s=z_sP_z$, ${\cal C}^{-1}_{\rm evo}$ and ${\cal C}^{-1}$ are the inverses of the DGLAP evolution kernel and the hybrid-scheme matching kernel, respectively. The detailed expressions for these kernels up to the NNLL accuracy are given in appendix~\ref{sec: matching}.

\begin{figure}[t!]
	\centering
    \vspace{-30pt}
	\includegraphics[width=0.58\textwidth]{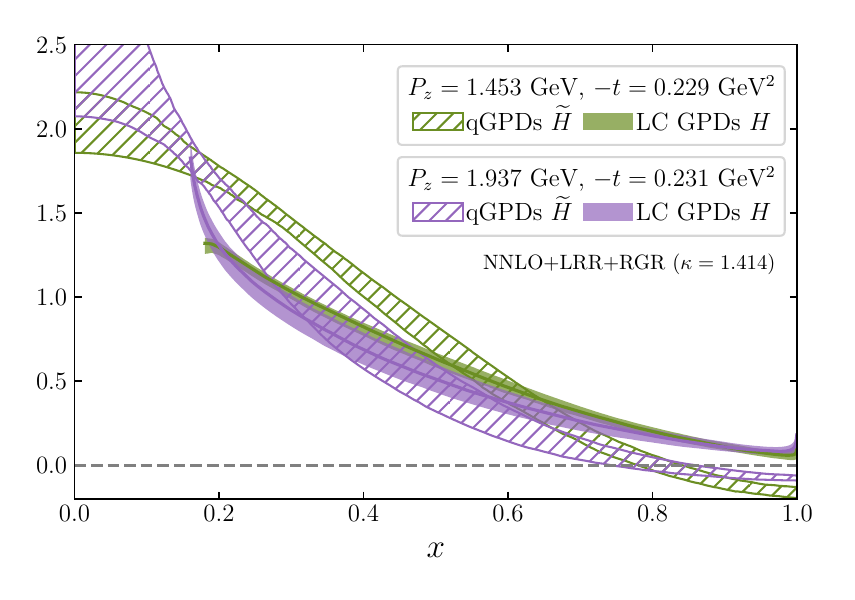}
	\caption{\label{fig: GPDs pz}
    The $x$ dependence of the qGPDs and light-cone (LC) GPDs, solved with NNLO+LRR+RGR ($\kappa=1.414$) perturbative matching, is shown. It contains the results from two datasets with different values of $P_z$ but similar values of $-t$.}
\end{figure}

Since the qGPDs vanish at the positive and negative infinities, the integration range is finite in practice. To obtain numerical results on the light-cone GPDs, we can discretize the integration variables and replace the integrals with sums, as shown in the second line of eq.~(\ref{eq: perturbative matching}). This procedure can reproduce the exact numerical integration well~\cite{Gao:2021dbh}. 
In figure~\ref{fig: GPDs pz}, we demonstrate both qGPD and light-cone GPD results obtained from NNLO+LRR+RGR ($\kappa=1.414$) perturbative matching with three-loop DGLAP evolution~\cite{Curci:1980uw, Moch:2004pa}. It shows the results from two datasets with different values of $P_z$ but similar values of $-t$. As seen from this figure, there is a significant difference in the $x$ dependence between the qGPDs and light-cone GPDs, especially at lower $P_z$. The effects of the matching are most significant at small and large values of $x$. Additionally, the $P_z$ dependence of the distributions is significantly corrected after matching, as expected. 
Since the strong coupling constant becomes ill-defined at low scales in RGR, causing the perturbation theory to break down, we can only calculate the results down to some minimum value of $x$.

Since a larger $P_z$ is favored for better control of power accuracy, and a $P_z\to\infty$ extrapolation on a single lattice ensemble cannot be rigorously done without controlling the discretization effects, we will focus on the case with the largest momentum $P_z=1.937$ GeV in the following discussion. We will explore the dependence of the light-cone GPDs on the matching accuracy, renormalization scale, and physical variables.

\subsection{The dependence of the light-cone GPDs on the perturbative accuracy of the matching and the renormalization scale}

In this subsection, we will examine the sensitivity of pion light-cone GPDs to both perturbative matching accuracy and renormalization scale, aiming to determine the reliability of our results.

\begin{figure}[t!]
	\centering
    \vspace{-30pt}
	\includegraphics[width=0.6\textwidth]{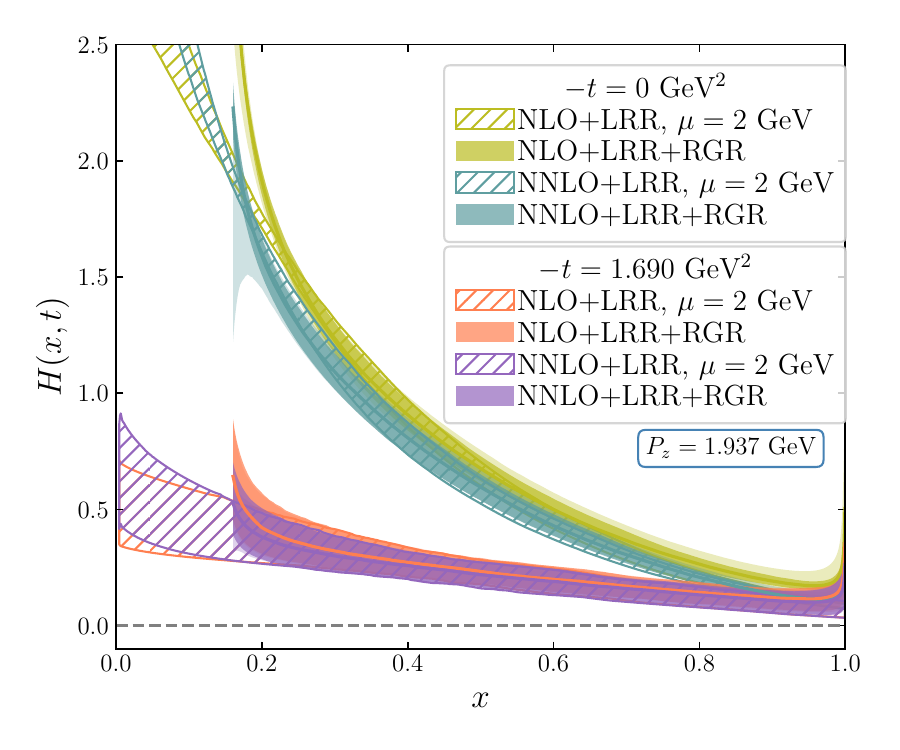}
	\caption{\label{fig: GPDs order pz4} The $x$ dependence of the valence light-cone GPDs for $P_z=1.937$ GeV is presented, showing results obtained with various perturbative matching choices and renormalization scales. The line-filled bands correspond to the results without RGR, while the solid-filled bands correspond to the results with RGR. For the latter, the darker-solid-filled bands represent the statistical uncertainty with $\kappa=1.414$, while the lighter-solid-filled bands contain additional systematic uncertainty with varying $\kappa$ from 1 to 2.}
\end{figure}

In figure~\ref{fig: GPDs order pz4}, we present the light-cone GPD results for the largest momentum $P_z=1.937$ GeV, obtained using NLO and NNLO perturbative matching with LRR, both with or without RGR.
The line-filled bands represent the results obtained without RGR at $\mu=2$ GeV. The solid-filled bands show the results obtained with RGR, considering the variation of $\kappa$. The darker-solid-filled bands display the statistical errors with $\kappa=1.414$ only, while the lighter-solid-filled bands also include additional systematic errors associated with scale variation from $\kappa=1$ to 2. The scale variation of the results is more significant at small
and large $x$. We have selected the data with the smallest and largest momentum transfers for a clearer comparison. The NNLO results exhibit smaller scale variation, particularly for data with larger $-t$ values. 
Moreover, the results with and without RGR are almost identical in the intermediate $x$ region but show significant differences at small $x$, as expected. As discussed above, due to the breakdown of perturbation theory, our light-cone GPD results obtained with RGR are only reliable for $x \gtrsim 0.2$. The significant changes in the vicinity of $x=1$ in RGR can be attributed to two factors. First, the quasi-GPD after inverse matching still has a non-vanishing tail at $x\ge1$ due to the power corrections, which propagates to the final result after DGLAP evolution. Besides, in this region the threshold logarithms in the matching and evolution kernels are important~\cite{Gao:2021hxl, Ji:2023pba,Ji:2024hit}. However, threshold resummation is beyond the scope of this work. Nevertheless, according to ref.~\cite{Ji:2024hit} both RGR and threshold resummations have little impact for $x<0.8$ at $P_z=1.937$ GeV, so we can focus on our results in the region $0.2\lesssim x \lesssim  0.8$ where the systematics is under control at current perturbative accuracy.
Finally, we note that the results between the NLO and NNLO display convergence, especially for the cases with larger $-t$ value.

\begin{figure}[t!]
	\centering
    \vspace{-30pt}
    \includegraphics[width=0.6\textwidth]{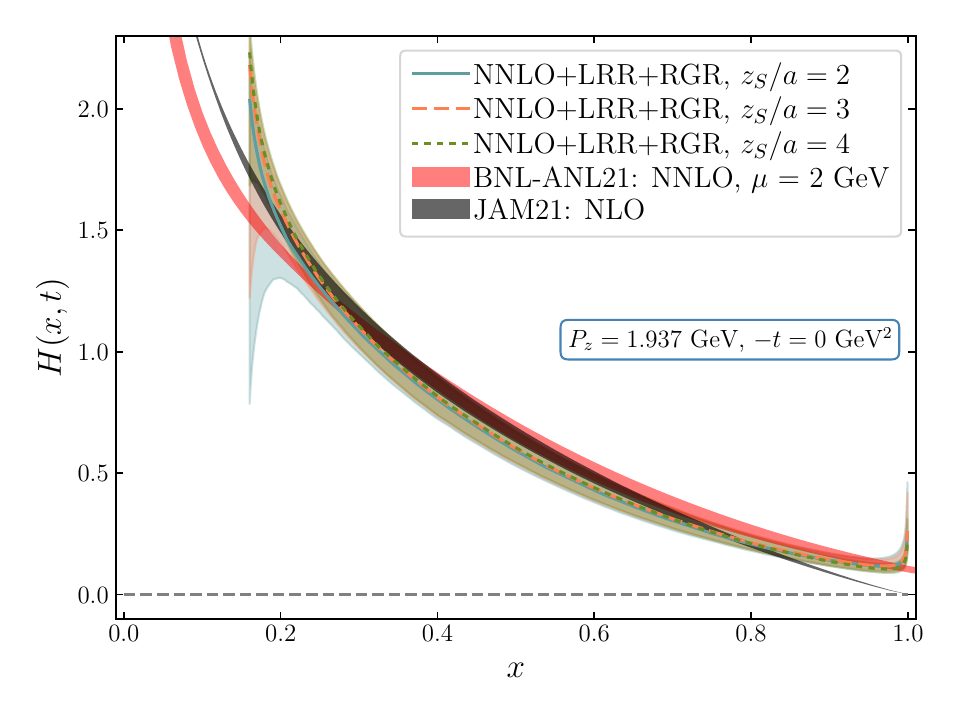}
	\caption{\label{fig: GPDs zS pz4} The dependence of the valence light-cone GPDs on $z_s$ is illustrated at $P_z = 1.937$ GeV and $-t = 0$ GeV$^2$. The black band displays the NLO global analyses results~\cite{Barry_2021}. The red band represents the valence pion PDFs obtained with NNLO matching in a hybrid scheme, excluding LRR and RGR~\cite{Gao:2021dbh}.}
\end{figure}

If one uses the hybrid scheme to perform the renormalization, the resulting light-cone GPDs may depend on the choice of $z_s$. Therefore, it is important to check the dependence of our results on $z_s$. We compare the valence light-cone GPD results with three different values of $z_s$ for $P_z=1.937$ GeV in figure~\ref{fig: GPDs zS pz4}. As a demonstration, we show our results for $-t=0$ GeV$^2$, obtained using NNLO+LRR+RGR. The lines represent the central values obtained with $\kappa=1.414$, while the bands encompass both statistical and systematic errors. This convention for the error bands is used in the subsequent figures as well. We can see that the final valence light-cone GPD results have little dependence on $z_s$ for $x>0.2$. At $0.2<x<0.3$, the results at $z_s=2a$ show a slight deviation from the other two cases, which is probably caused by the discretization effects as $z_s\sim a$. The situations are similar for other values of $-t$. Thus, our predictions in the region $0.2\lesssim x\lesssim 0.8$ are not sensitive to the choice of $z_s$.
Considering that the value of $\bar{m}_0$ at $z/a=3$ exhibits the minimal dependence on the perturbative order and renormalization scale, we adopt the corresponding position $z_s/a=3$ for subsequent analyses.

Furthermore, the zero momentum transfer case corresponds to the PDF results, allowing for comparison with previous PDF calculations. Specifically, we compare our findings with the global analyses at NLO accuracy (JAM21)~\cite{Barry_2021} and our previous results using the NNLO matching (without LRR or RGR) within the same lattice framework (BNL-ANL21)~\cite{Gao:2021dbh}. Our results agree well with the global fit results for $0.2\lesssim x \lesssim 0.8$. Regarding the comparison with BNL-ANL21, the results show good agreement in the range $0.3\lesssim x \lesssim 0.7$, while the differences observed in the small and large $x$ regions are expected due to contributions from both LRR and RGR.

\subsection{The $t$-dependence of the light-cone GPDs}

\begin{figure}[t!]
	\centering
    \vspace{-30pt}
    \includegraphics[width=0.6\textwidth]{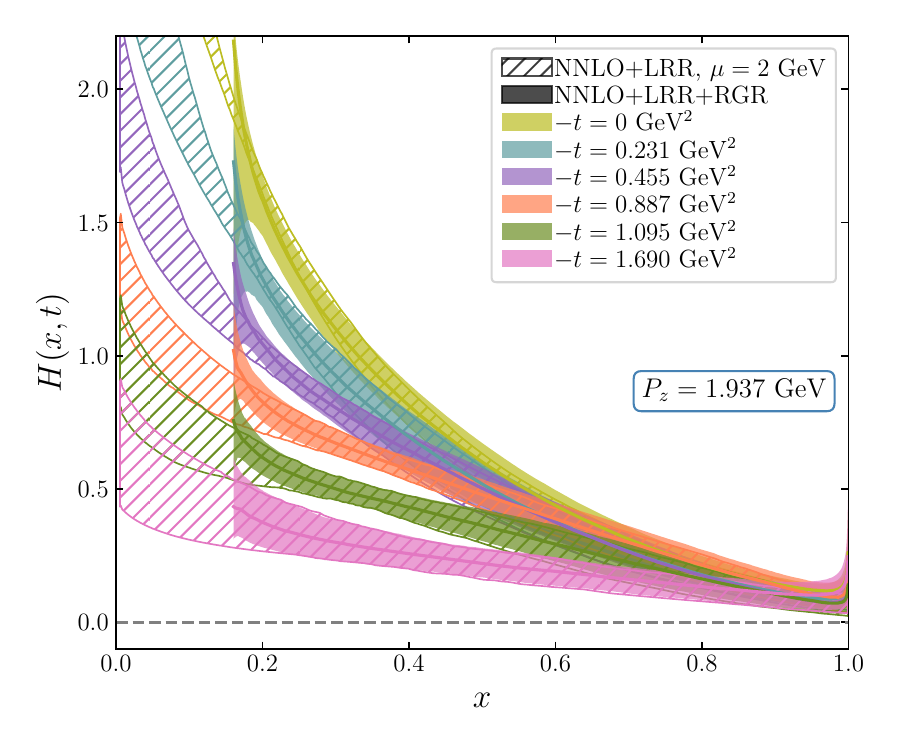}
	\caption{\label{fig: GPDs Q2} The valence pion light-cone GPDs as a function of $x$ for all available values of $-t$. The solid-filled bands represent results obtained with NNLO+LRR+RGR, and the width of the bands indicates the total uncertainty of the results, including scale variation. The line-filled bands represent NNLO+LRR results for $\mu=2$ GeV.}
\end{figure}

Having analyzed the systematic effects in the determination of the valence pion light-cone GPDs at some representative values of $-t$, we can now study the entire $t$ dependence of the light-cone GPDs in more detail.
In figure~\ref{fig: GPDs Q2}, we present the valence light-cone GPDs as a function of the Bjorken-$x$ across all the available values of $-t$. 
As expected from the previous discussion, for $x<0.2$, there are significant differences in all results between the cases with (solid-filled bands) and without (line-filled bands) RGR.
The distributions exhibit significant dependence on $-t$ and $x$. Firstly, the light-cone GPDs $H(x,\xi=0,t)$ decrease as $-t$ increases for a fixed $x$. Secondly, the fall-off of the GPDs along $x$ is notably reduced at larger values of $-t$. 
Similar behavior of the pion GPDs was also found in ref.~\cite{Lin:2023gxz}, where a lattice spacing of $a=0.09$ fm and a symmetric frame were utilized.

\begin{figure}[t!]
	\centering
    \vspace{-30pt}
	\includegraphics[width=0.48\textwidth]{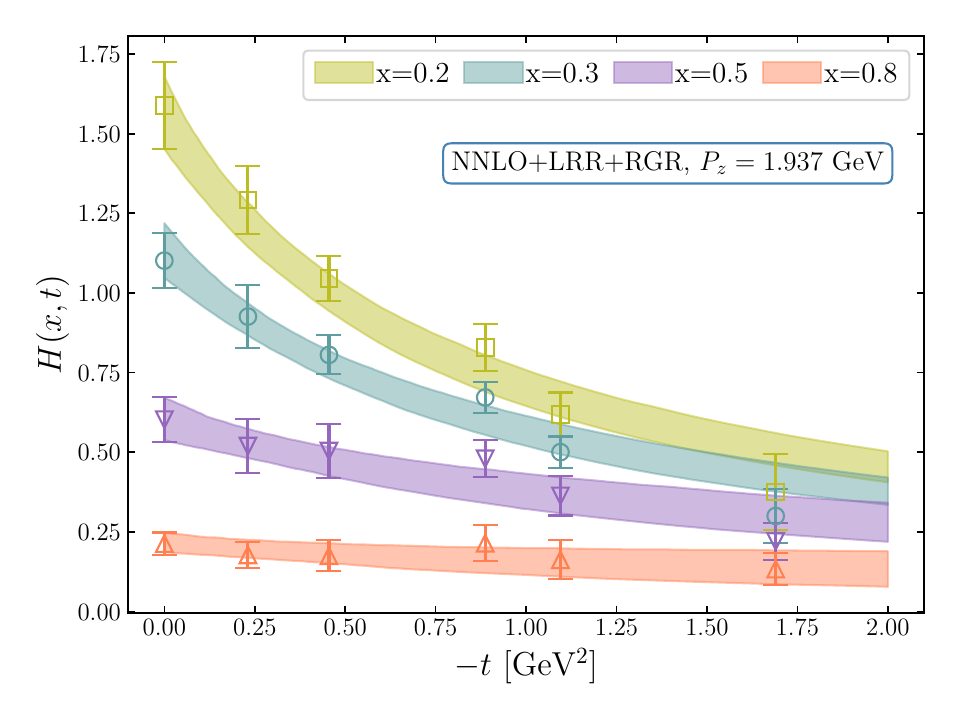}
    \includegraphics[width=0.48\textwidth]{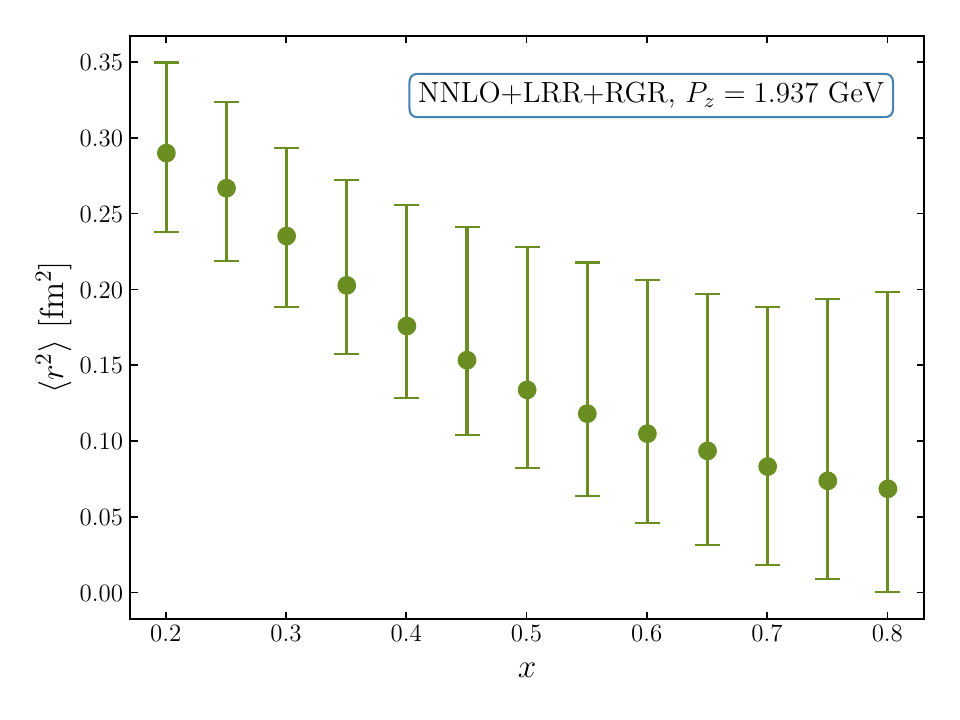}
	\caption{\label{fig: GPDs x} Left: The $t$-dependence of the valence light-cone GPDs for various values of $x$. Right: The pion effective radius as a function of the Bjorken-$x$.}
\end{figure}

To further explore the $t$ and $x$ dependence of the pion valence light-cone GPDs, we would like to parametrize our results with a suitable ansatz. Based on past experience with the pion form factor~\cite{Gao:2021xsm}, we employ the monopole form
\begin{equation}
    H(x, t) = \frac{H(x, 0)}{1-t/M^2(x)},
\end{equation}
where $H(x, 0)$ represents the GPDs at zero momentum transfer, equivalent to the PDFs, and $M(x)$ is a $x$-dependent monopole mass parameter, which is related to the pion effective radius as $\langle r^2 (x) \rangle=6/M^2(x)$, in analogy to the usual pion charge radius. This fit ansatz works very well, as demonstrated by the bands in the left panel of figure~\ref{fig: GPDs x}. The left panel displays the $t$ dependence of the light-cone GPDs for four selected values of $x\in[0.2,0.3,0.5,0.8]$. The lattice results are shown as the data points, and the corresponding monopole fit results are displayed as the bands. We can observe that the $t$ dependence of the valence light-cone GPDs becomes milder as $x$ increases. For $x=0.8$, almost no $t$ dependence can be discerned within the estimated errors. The results of the pion effective radius are shown in the right panel of figure~\ref{fig: GPDs x}. The asymmetric error bars reflect the underlying skewed distribution of the effective radius sample results. Given that radius cannot be negative, we use the median as the central value, with the lower and upper uncertainties defined by the 16th and 84th percentiles of the sample distribution, respectively.
The effective radius clearly decreases with increasing $x$.
This means that when quarks have higher momentum fractions, they are likely to be confined to a smaller spatial region in the transverse plane. This confinement results in narrower distributions in position space, potentially reducing the effective radius.
Interestingly, the effective radius for $x=0.2$ is comparable to the pion charge radius obtained from the pion electromagnetic form factor, specifically $\langle r_{\pi}^2 \rangle =0.313(27)$ fm$^2$, as calculated in ref.~\cite{Gao:2021xsm} using the same pion mass.

\begin{figure}[t!]
	\centering
    \vspace{-30pt}
	\includegraphics[width=0.7\textwidth]{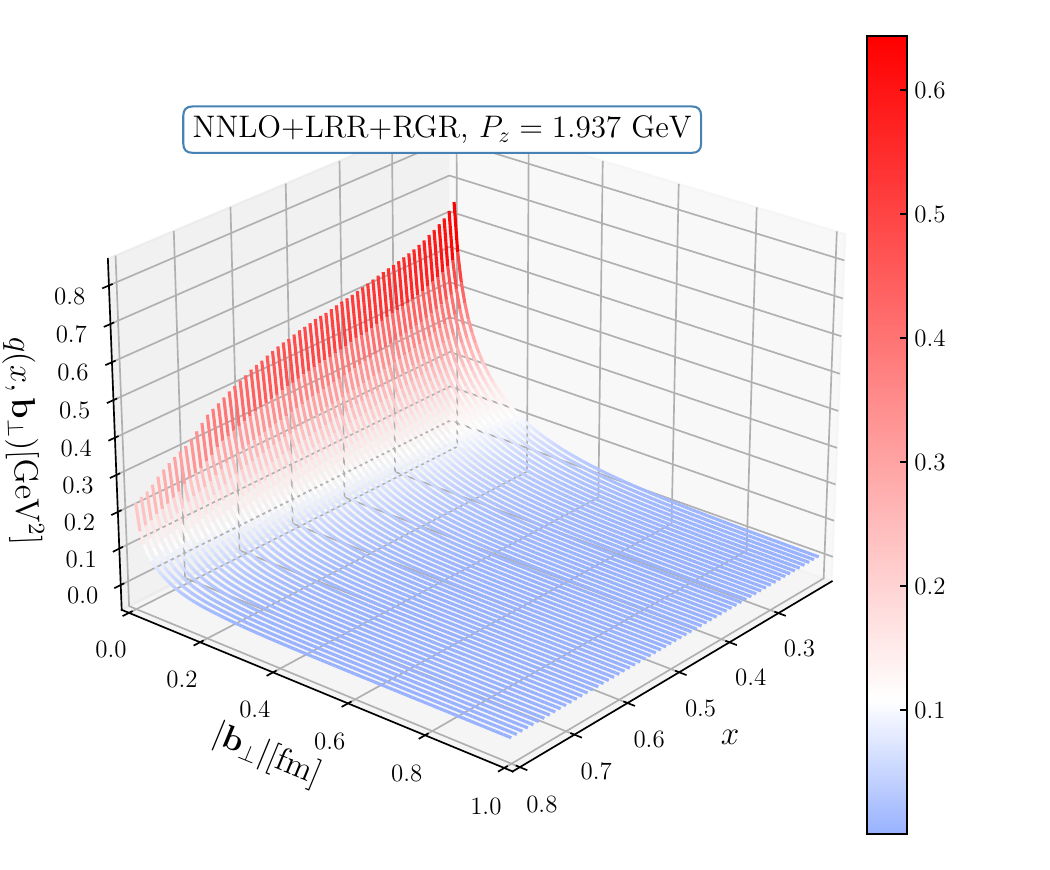}
	\caption{\label{fig: IPD} The valence IPDs are shown as functions of the Bjorken-$x$ and the impact parameter $\mathbf{b}_\perp$.}
\end{figure}

Moreover, the Fourier transform of the valence light-cone GPDs at zero skewness, with respect to the transverse components of the momentum transfer, defines the valence impact-parameter-space parton distributions (IPDs): 
\begin{equation}
    q(x, \mathbf{b}_\perp) = \int \frac{{\rm d}^2 \mathbf{\Delta}_\perp}{(2\pi)^2} H(x, \mathbf{\Delta}^2_\perp) e^{i\mathbf{b}_\perp\cdot\mathbf{\Delta}_\perp},
\end{equation}
where $\mathbf{b}_\perp$ is the impact parameter. The IPDs describe the probability density of finding a parton with momentum fraction $x$ at a specific transverse distance $\mathbf{b}_\perp$ from the center of the transverse momentum (CoTM). They can offer a comprehensive view of the parton distributions in both momentum and coordinate spaces within the hadron. Using our monopole fit results of $H(x,t)$, we can perform a Fourier transform in $\mathbf{\Delta}_\perp$ to obtain the valence IPDs. Figure~\ref{fig: IPD} shows the results with varying $x$ from 0.2 to 0.8 and $\mathbf{b}_\perp$ from 0 to 1 fm. We can find that when the partons carry a larger momentum fraction $x$, the distributions are more localized, which results in a more concentrated distribution in the impact-parameter space. Conversely, as $x$ decreases, the distributions become broader in $\mathbf{b}_\perp$, indicating a more diffuse spatial distribution of lower-momentum quarks.

\begin{figure}[t!]
	\centering
    \vspace{-30pt}
	\includegraphics[width=0.7\textwidth]{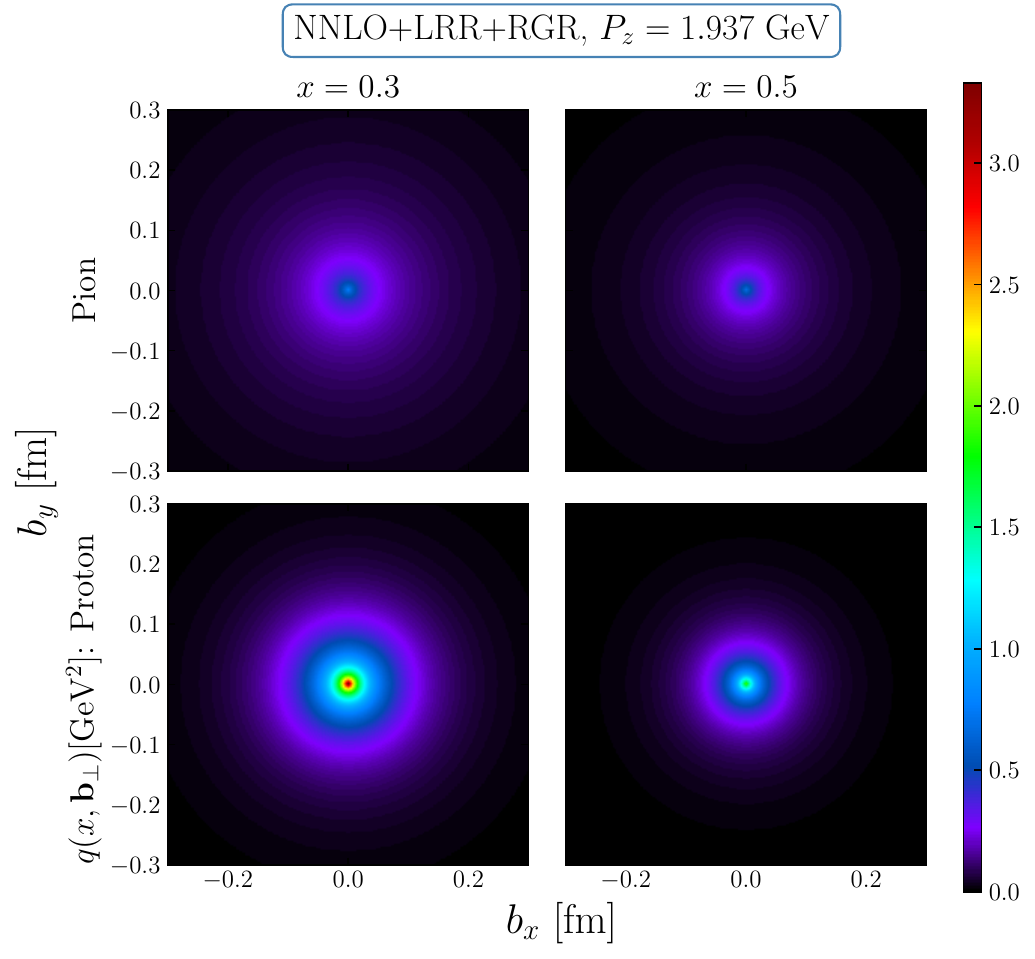} \\
	\caption{\label{fig: bT} The two-dimensional distributions show a comparison of the IPDs between the pion and proton \cite{Cichy:2023dgk} at $x$=0.3 and 0.5.}
\end{figure}

It is interesting to compare the pion valence IPDs with those of the proton, as shown in figure~\ref{fig: bT}.
We display the comparison in a two-dimensional distribution at $x=0.3$ and 0.5. The light-cone proton GPDs data are from ref.~\cite{Cichy:2023dgk}. We perform a dipole fit on the proton GPDs and then get the corresponding IPDs through the Fourier transform. For both the pion and proton, the distributions become more concentrated at larger $x$, corresponding to smaller effective radii, which is consistent with our findings from the right panel of figure~\ref{fig: GPDs x}. Comparison of the proton and pion results at fixed $x$ reveals that the proton exhibits a broader distribution of valence quarks than the pion at both $x=0.3$ and 0.5. To some extent, this observation relates to the fact that the charge radius of the proton is larger than that of the pion. Figure~\ref{fig: bT} also indicates that the valence quark distributions within the pion are somewhat less peaked at the very center compared to those within the proton.

\newpage
\section{\label{sec: conclusion} Conclusions}

We study the pion valence light-cone GPDs using the LaMET approach at a lattice spacing of 0.04 fm, which is more than two times smaller than the lattice spacing used in other studies of pion GPDs ($a=0.09$ fm)~\cite{Lin:2023gxz}. In this work, we utilize the Lorentz-invariant
definition of the GPDs, enabling us to perform calculations in the non-Breit frame. This approach enables simultaneous calculation of multiple momentum transfers and thus significantly reduces the computational costs. In this work, we perform calculations within both the Breit and non-Breit frames. By comparing the amplitude results of these two frames, we demonstrated once again the validity and efficacy of the Lorentz-invariant approach.
To obtain the pion valence light-cone GPDs from the qGPDs, we use NNLO hybrid-scheme matching along with leading renormalon resummation and renormalization group resummation. The use of leading renormalon resummation significantly reduces the perturbative uncertainty in the hybrid renormalization scheme. Furthermore, the use of renormalization group resummation allows us to determine the range of Bjorken-$x$ for which the LaMET approach is reliable for a specific momentum value $P_z$. Our results at zero momentum transfer, specifically the valence light-cone PDF results, are in good agreement with the global analyses~\cite{Barry_2021} within the range $0.2\lesssim x \lesssim 0.8$. The $P_z$ dependence of the final light-cone GPD results is significantly reduced compared to the qGPDs, which indicates the effectiveness of the perturbative matching framework. Our lattice calculations provide a detailed three-dimensional imaging of the pion structure, which can be better visualized in terms of impact-parameter-dependent distribution. From these findings, we observe that the effective transverse size of the pion reduces as $x$ increases, a pattern also observed in the proton~\cite{Cichy:2023dgk}.
Additionally, the effective size of the pion is consistently smaller than that of the proton for the considered $x$ region. For future studies, several improvements would enhance our understanding: First, increasing statistics and extending $t_s$ for larger momenta would enable reliable 3-state fit results. This would allow for a more thorough investigation of excited state contamination. Second, while we acknowledge uncertainties in our preditions for $x<0.2$ and $x>0.8$ regions, incorporating threshold resummation in perturbative matching procedure could yield more precise results in the large-$x$ region. Finally, studying much larger momentum values would provide better insights into higher-twist contributions.

\acknowledgments
We thank Shohini Bhattacharya, Yushan Su, and Rui Zhang for their valuable communications.

This material is based upon work supported by the U.S. Department of Energy, Office of Science, Office of Nuclear Physics through Contract Nos. DE-SC0012704, DE-AC02-06CH11357, and within the frameworks of Scientific Discovery through Advanced Computing (SciDAC) award Fundamental Nuclear Physics at the Exascale and Beyond, and under the umbrella of the Quark-Gluon Tomography (QGT) Topical Collaboration with Award DE-SC0023646. HTD is supported by the NSFC under Grant Nos. 12325508, 12293060, 12293064, and the National Key Research and Development Program of China under Grant No. 2022YFA1604900. QS is partially supported by Laboratory Directed Research and Development (LDRD No. 23-051) of BNL and RIKEN-BNL Research Center. SS is supported by the National Science Foundation under CAREER Award PHY-1847893. YZ is partially supported by the 2023 Physical Sciences and Engineering (PSE) Early Investigator Named Award program at Argonne National Laboratory.

This research used awards of computer time provided by the U.S. Department of Energy’s INCITE and ALCC programs at the Argonne and the Oak Ridge Leadership Computing Facilities. The Argonne Leadership Computing Facility at Argonne National Laboratory is supported by the Office of Science of the U.S. DOE under Contract No. DE-AC02-06CH11357. 
The Oak Ridge Leadership Computing Facility at Oak Ridge National Laboratory is supported by the Office of Science of the U.S. DOE under Contract No. DE-AC05-00OR22725. 
This research also used the Delta advanced computing and data resource which is supported by the National Science Foundation (award OAC 2005572) and the State of Illinois. Delta is a joint effort of the University of Illinois Urbana-Champaign and its National Center for Supercomputing Applications.
Computations for this work were carried out in part on facilities of the USQCD Collaboration, funded by the Office of Science of the U.S. Department of Energy.

\appendix

\section{\label{sec: fit results}Fit results of the two-point and three-point correlation functions}

\begin{figure}[t!]
	\centering
    \vspace{-20pt}
	\includegraphics[width=0.5\textwidth]{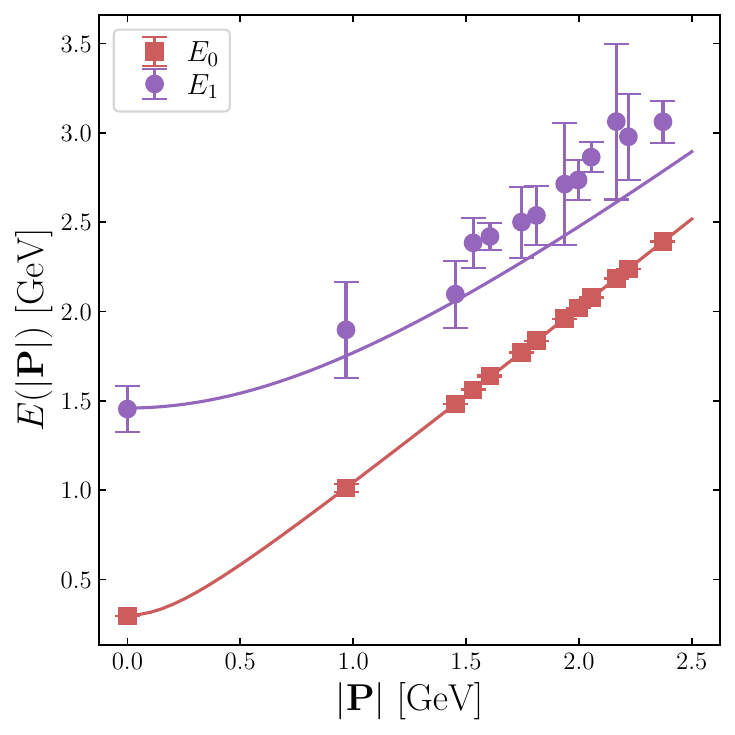}
	\caption{\label{fig: DR} The results for the ground state energy $E_0$ and the first excited state energy $E_1$, extracted from the two-point functions, are shown. The solid lines represent the results calculated using the dispersion relation $E_n=\sqrt{m^2_n+|\textbf{P}|^2}$ with $m_0$ = 0.3 GeV and $m_1$ = 1.46 GeV.}
\end{figure}

\begin{figure}[t!]
	\centering
	\includegraphics[width=0.47\textwidth]{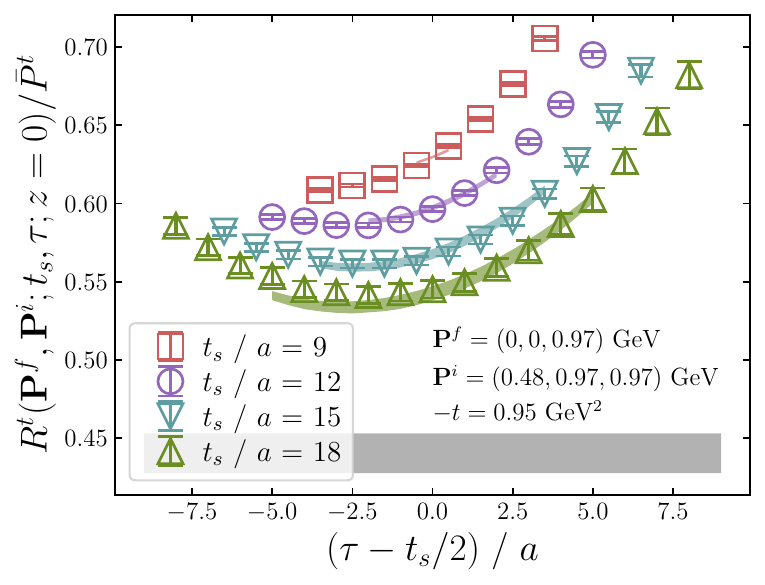}
	\includegraphics[width=0.47\textwidth]{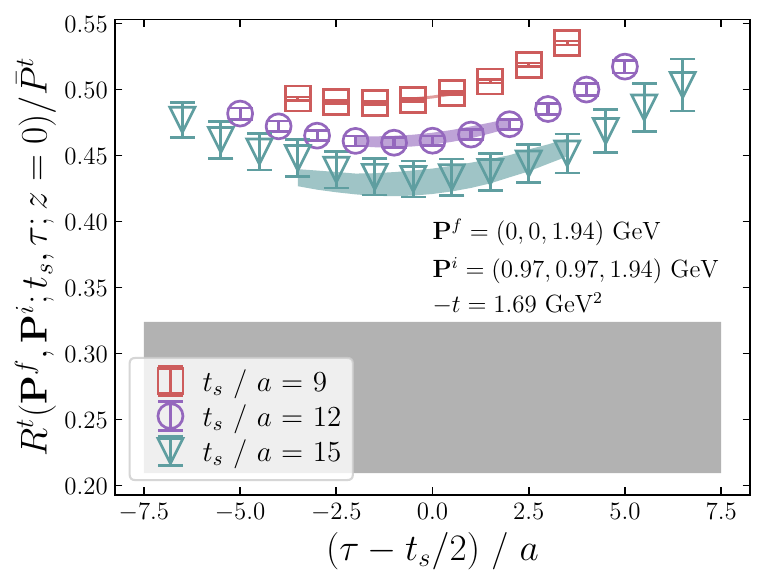} 
	\caption{\label{fig: ratio} Two examples of the ratio $R^t(z=0)/\bar{P}^t$ are shown. Left: $P_z$ = 0.97 GeV and $-t$ = 0.95 GeV$^2$, right: $P_z$ = 1.94 GeV and $-t$ = 1.69 GeV$^2$. The data points represent the lattice results, and the bands with corresponding colors are the fit results of the lattice data. The grey bands show the results of the bare matrix elements.}
\end{figure} 

In this appendix, we present the detailed analyses of the two-point and three-point correlation functions. As mentioned in the main text, by employing the spectral decomposition formula along with 2-state or 3-state fits of the pion two-point correlators~\cite{Gao:2020ito, Gao:2021dbh, Gao:2021xsm, Gao:2022iex, Gao:2022vyh},
we can extract energy values of both the ground and excited states of the pion, as well as the corresponding amplitudes.
We show the results of the ground and the first excited state energy for different momenta in figure~\ref{fig: DR}. We can see that the fit results for $E_0$ agree well with the dispersion relation shown by the solid lines, and in most cases, the results for $E_1$ also agree with them within errors.

To obtain the matrix elements, we consider the ratio $R^{\mu}$ given by eq.~(\ref{eq: ratio}) and perform 2-state or 3-state fits.
Here, we use the energy levels and amplitudes obtained from the corresponding fits of the two-point functions.
In figure~\ref{fig: ratio}, we select two specific cases from the data of $z=0$ and $\mu=t$ to show the results of the ratio as $R^t/\bar{P}^t$. The left one belongs to the case of $P_z$ = 0.97 GeV and $-t$ = 0.95 GeV$^2$, while the right one corresponds to the case of $P_z$ = 1.94 GeV and $-t$ = 1.69 GeV$^2$, that is, the largest momentum transfer in this work. The lattice results are shown as the data points with different colors representing the different time separations. The bands with corresponding colors denote the two-state fit results of the lattice data, providing a good description of the lattice results. And the grey bands are the extrapolated results of the ratio, that is, the bare matrix elements. 

\begin{figure}[t!]
	\centering
    \vspace{-25pt}
	\includegraphics[width=0.52\textwidth]{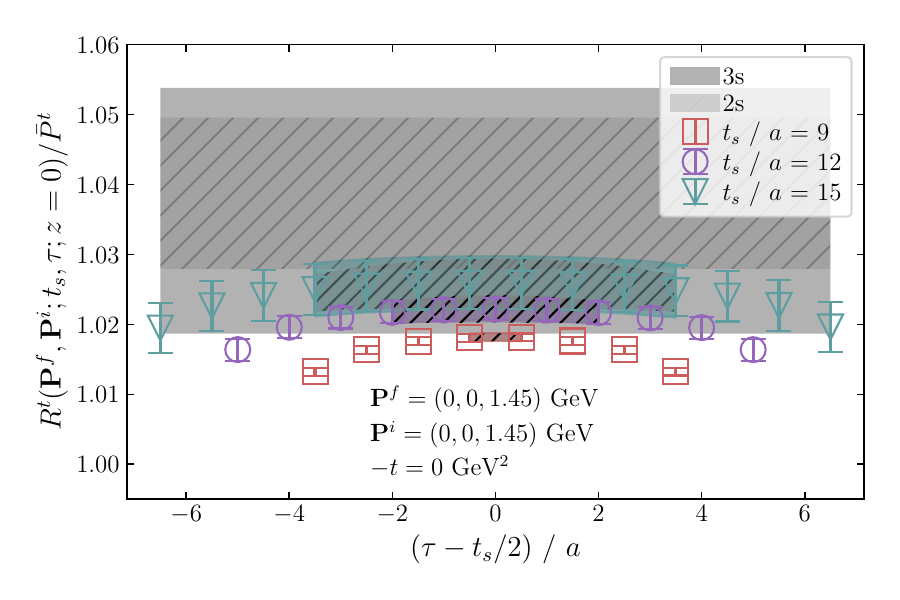}
	\includegraphics[width=0.46\textwidth]{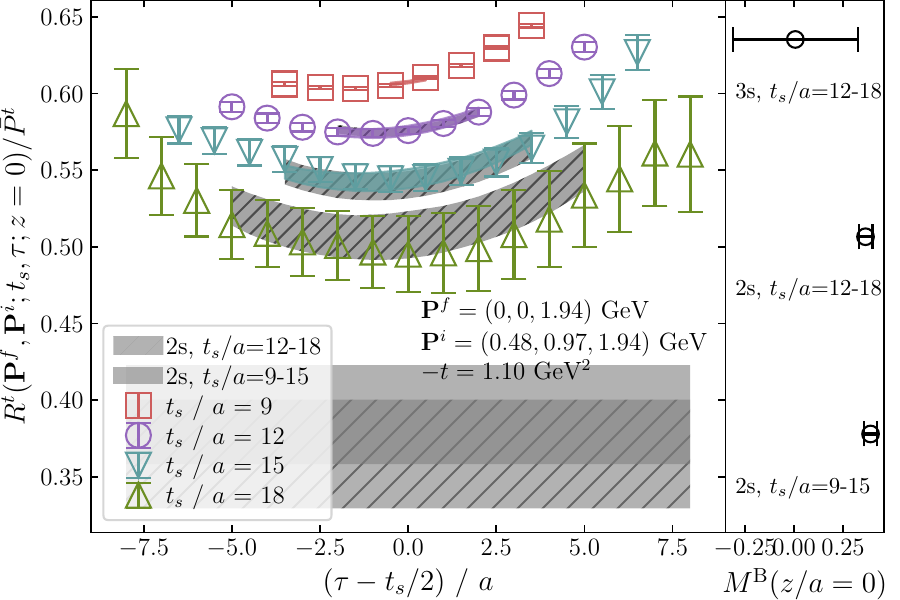} 
	\caption{\label{fig: ratio 3s} Two examples of the ratio $R^t(z=0)/\bar{P}^t$ used to study the effectiveness of a 2-state fit with $t_s/a\in[9, 12, 15]$ for large momentum. Left: $P_z$ = 1.45 GeV at $-t$ = 0 GeV$^2$, right: $P_z$ = 1.94 GeV at $-t$ = 1.10 GeV$^2$. The data points and the bands have the same meaning as in figure~\ref{fig: ratio}.}
\end{figure} 

To address excited state contamination for the largest two momenta, we conducted additional analyses of $R^t(z=0)/\bar{P}^t$ using multiple approaches. Figure~\ref{fig: ratio 3s} presents two representative cases demonstrating the reliability of a 2-state fit with the largest $t_s$ being 0.6 fm.
For $P_z$ = 1.45 GeV at $-t$ = 0 GeV$^2$ (left panel), we compare the results from 2-state (line-filled bands) and 3-state fits (solid-filled bands) using $t_s\in[0.36, 0.48, 0.6]$ fm ($t_s/a\in[9, 12, 15]$). The agreement between these two fits within uncertainties supports the adequacy of 2-state fit at large momentum.
For $P_z$ = 1.94 GeV at $-t$ = 1.10 GeV$^2$ (right panel), we extended our analysis to larger source-sink separations. We compare 2-state fits using two ranges of $t_s\in [0.36, 0.48, 0.6]$ fm (solid-filled bands, labeled as $t_s/a=9-15$) and $t_s\in[0.48, 0.6, 0.72]$ fm (line-filled bands, labeled as $t_s/a=12-18$). The agreement between fits across different $t_s$ ranges supports the reliability of the ground state extraction. Furthermore, the right sub-figure of the right panel shows the bare matrix element results at $z/a=0$. When we attempted 3-state fit with the extended $t_s$ range, the combination of high momentum and large momentum transfer led to prohibitively large uncertainties, making these fits statistically unreliable. The additional analyses above demonstrate that a two-state fit with the largest $t_s$ being 0.6 fm can provide stable and reliable results, even at large momenta and momentum transfers.

\begin{figure}[t!]
	\centering	
    \vspace{-20pt}
    \includegraphics[width=0.55\textwidth]{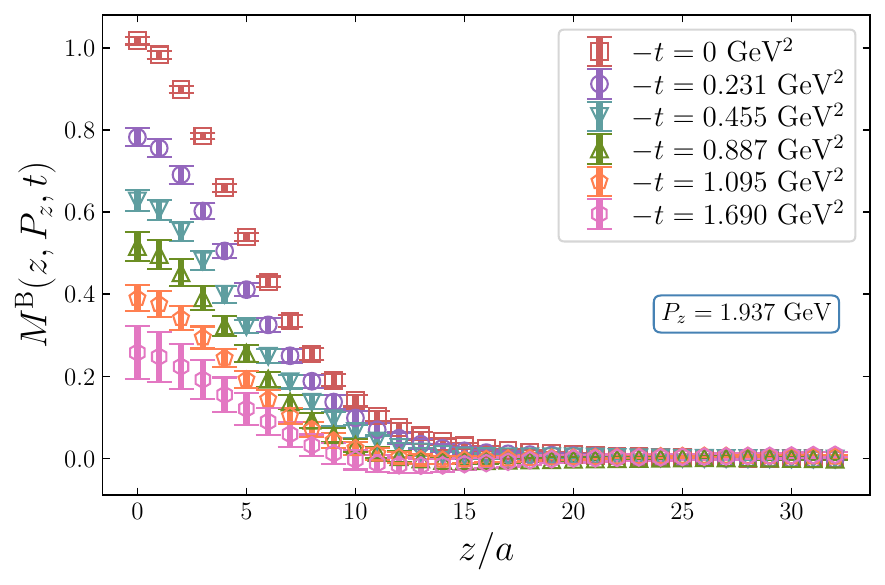}
	\caption{\label{fig: normME} The bare matrix element results for the largest momentum $|\mathbf{P}^f| = P_z = 1.937$ GeV are shown as a function of $z$.}
\end{figure}

We collect all the bare matrix element results of the largest momentum in figure~\ref{fig: normME}. It is evident that the results display a decreasing trend with increasing $z$ and also as the momentum transfer increases, as expected.

\newpage
\section{\label{sec: kappa}Detailed analysis of scale variation parameter selection}

\begin{figure}[t!]
    \centering
    \includegraphics[width=0.5\linewidth]{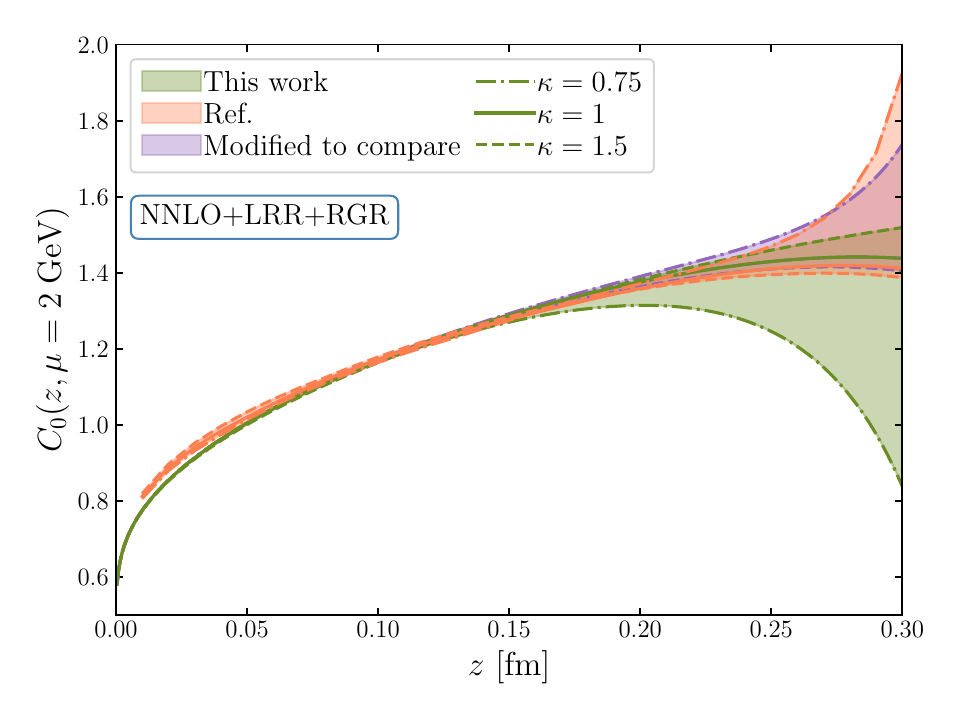} \hspace{-15pt}
    \includegraphics[width=0.5\linewidth]{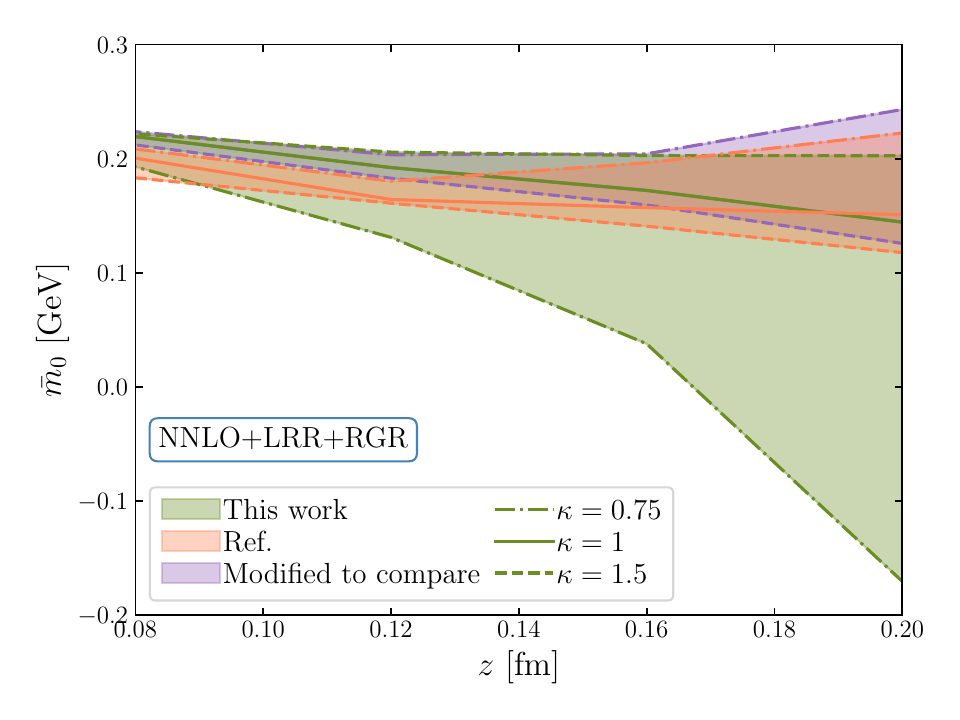}
    \caption{Comparison of $C_0(z, \mu=2 \text{ GeV})$ (left panel) and $\bar{m}_0$ (right panel) derived from NNLO+LRR+RGR perturbative calculations between our results and ref.~\cite{Zhang:2023bxs}.}
    \label{fig: C0 and m0}
\end{figure}

In an OPE, it is natural to choose $\kappa=1$ so that the apparent logarithms vanish in the Wilson coefficients. In practice the choice of $\kappa$ could depend on the observable, but it still should not be too different from $1$. After all, regardless of the exact choice, the key is to estimate the associated theory uncertainty by varying $\kappa$ in a given range, such as $\kappa\in [0.5, 2]$ that is typically used in collider phenomenology. In this work, the range of $\kappa$ are mainly decided by the behavior of $C_0$ and $\bar{m}_0$. Since the studies of $C_0$ and $\bar{m}_0$ at NNLO+LRR+RGR level are very limited, it is valuable to make a comparison with recent findings from ref.~\cite{Zhang:2023bxs}. In figure~\ref{fig: C0 and m0}, we compare our results for $C_0(z, \mu=2 \text{ GeV})$ (left panel) and $\bar{m}_0$ (right panel) derived from NNLO+LRR+RGR perturbative calculations with those from ref.~\cite{Zhang:2023bxs}. Our results are shown in green, while those from the reference are displayed in orange. The results for $\kappa\in[0.75, 1, 1.5]$ are represented with distinct line styles. 

We observe that at $\kappa=1$ our results and the results of ref.~\cite{Zhang:2023bxs} more or less agree. Both results exhibit larger scale variation when $\kappa < 1$. However, for large $z$ the variation with $\kappa$ is larger in our case. Furthermore, 
the result of ref.~\cite{Zhang:2023bxs} increases for $\kappa<1$, while our result decreases.
This discrepancy is due to the differences in how the scale variation parameter ($\kappa$) is fixed: we use a consistent choice of the scale variation parameter throughout the analysis. In  ref.~\cite{Zhang:2023bxs} the authors fix the scale variation parameter in the evolution factor before the asymptotic form to 1, while allowing variation in $\kappa$ in other parts of the expression. (More details about the approach used in the reference can be found in a PhD thesis~\cite{Zhang:2023tnc}.) If we modify our formula to match the procedure in ref.~\cite{Zhang:2023bxs}, the corresponding results (shown in purple in figure~\ref{fig: C0 and m0}) align reasonably well with those in ref.~\cite{Zhang:2023bxs}. 

\begin{figure}[t!]
    \centering	
    \vspace{-20pt}
    \includegraphics[width=0.6\linewidth]{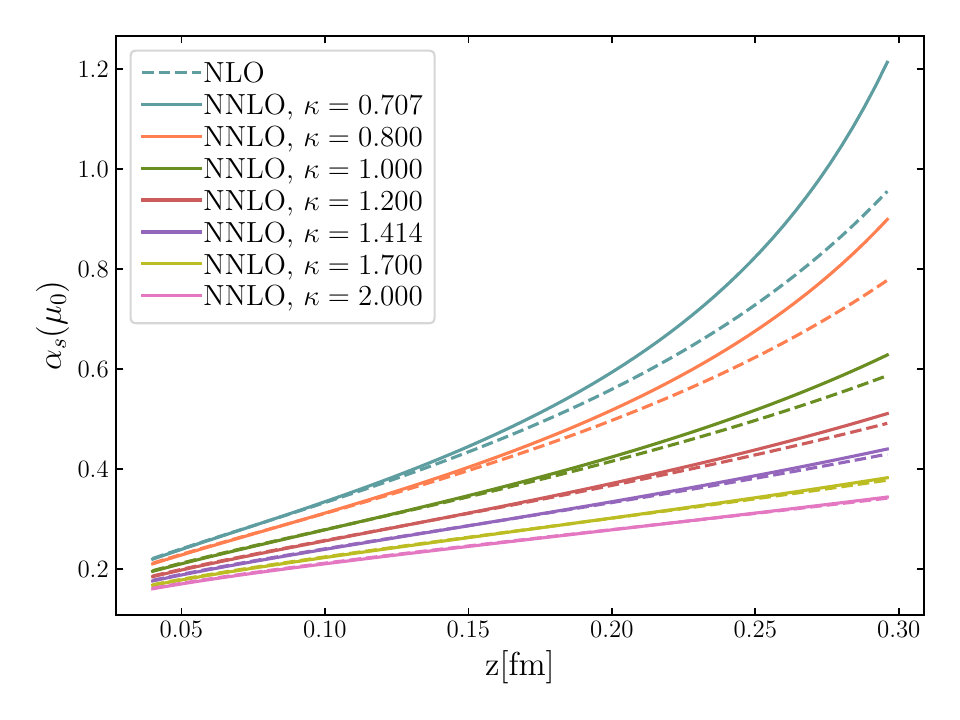}
    \caption{The results of running coupling constant derived with various values of $\kappa$ using NNLO(NLO)+LRR+RGR perturbative calculations. The solid lines correspond to the NNLO results, while the dashed lines correspond to the NLO results.}
    \label{fig: alpha allkappa}
\end{figure}
\begin{figure}[t!]
    \centering
    \includegraphics[width=0.8\linewidth]{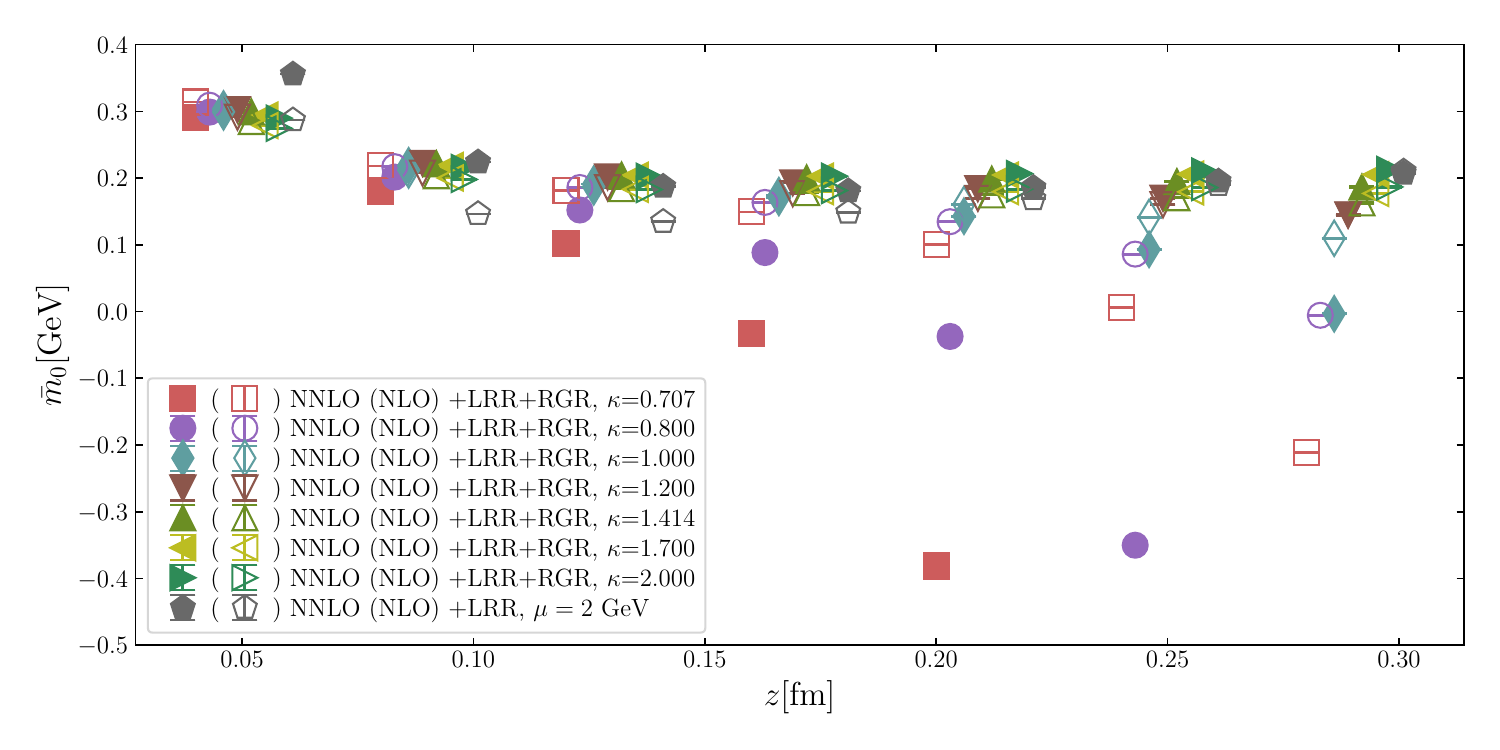}
    \caption{Same as figure~\ref{fig: m0}, but with a broader range of $\kappa$.}
    \label{fig: m0 allkappa}
\end{figure}

Regarding the significant scale dependence at lower values of $\kappa$ in our consistent scale variation coefficient approach, we explore a broad range of $\kappa$ to calculate $\alpha_s$ and $\bar{m}_0$ as shown in figures~\ref{fig: alpha allkappa} and \ref{fig: m0 allkappa}. Based on these results, at short distances where perturbation theory is applicable, they behave well as expected for $1<\kappa<2$, since $\alpha_s$ remains reasonably small and $\bar{m}_0$ remains constant. However, for lower $\kappa$ values, $\alpha_s$ becomes larger and $\bar{m}_0$ diverges under the NNLO+LRR+RGR framework. Consequently, we select $\kappa\in[1, 1.414, 2]$ as our range for scale variation, ensuring the validity of perturbative calculations while maintaining reasonable scale variation to estimate uncertainties.

\section{\label{sec: z dep of m0}Dependence of the results on $\bar{m}_0$ at different values of $z$}

\begin{figure}[t!]
    \centering	
    \vspace{-20pt}
    \includegraphics[width=0.5\linewidth]{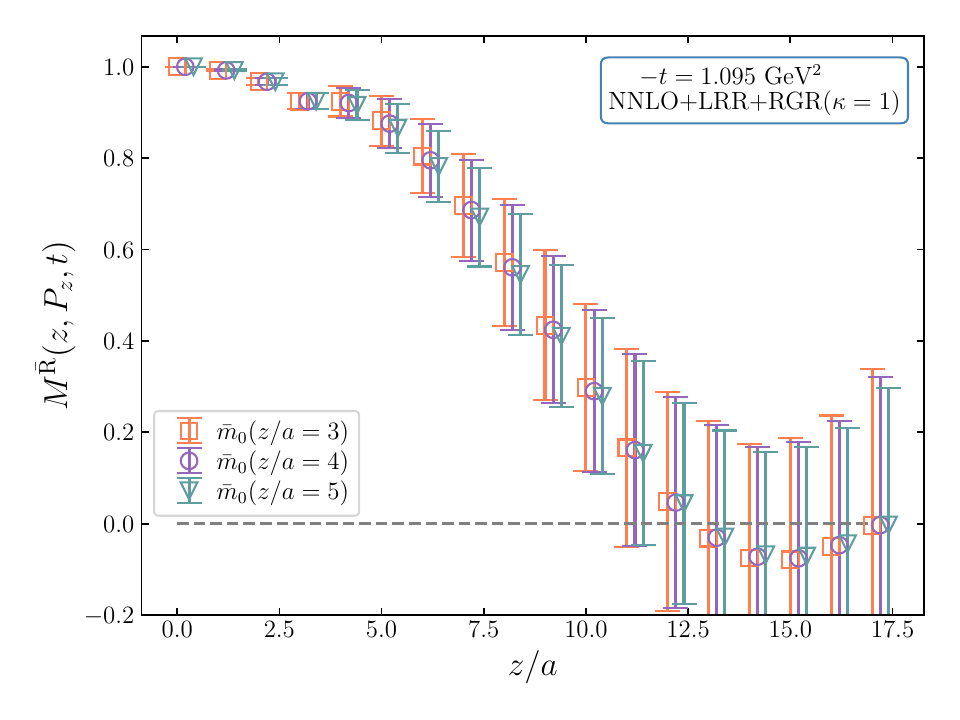} \hspace{-15pt}
    \includegraphics[width=0.5\linewidth]{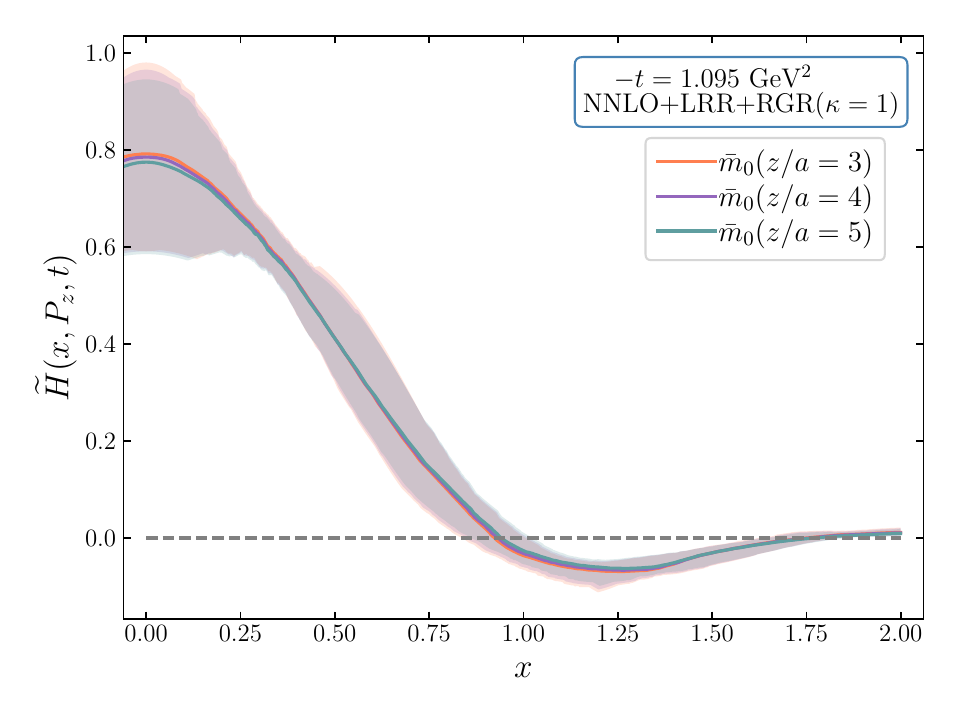}
    \caption{Comparison of the renormalized matrix element $M^{\bar{R}}$ (left panel) and quasi-GPD $\widetilde{H}$ (right panel) derived using $\bar{m}_0$ at various values of $z$ for the case of $-t=1.095$ GeV$^2$ with NNLO+LRR+RGR($\kappa=1$) perturbative matching.}
    \label{fig: m0 at diff z}
\end{figure}

To investigate the dependence of the results on $\bar{m}_0$ at different values of $z$, we consider the case of $-t=1.095$ GeV$^2$ with NNLO+LRR+RGR($\kappa=1$) perturbative matching as an example, selecting additional values of $\bar{m}_0$ at $z/a=4$ and 5 to do the renormalization. The renormalized matrix element $M^{\bar{R}}$ and the quasi-GPD, shown in figure~\ref{fig: m0 at diff z}, are then compared with those calculated using $\bar{m}_0$ at $z/a=3$. It can be found that variations in $\bar{m}_0$ across different $z$ values have minimal effect on the renormalized results, as the difference in $\bar{m}_0$ between $z/a=3$ and 5 ($\sim$ 0.05 GeV) is negligible compared to the value of $\delta m+\bar{m}_0$ ($\sim$ 0.93 GeV). Therefore, using $\bar{m}_0$ from different $z$ does not significantly impact our results on the quasi-GPD as can be seen from the right panel of figure~\ref{fig: m0 at diff z}.

\section{\label{sec: extrapolation}Extrapolation of the renormalized matrix elements}
\begin{figure}[t!]
    \centering	
    \vspace{-20pt}
    \includegraphics[width=0.9\linewidth]{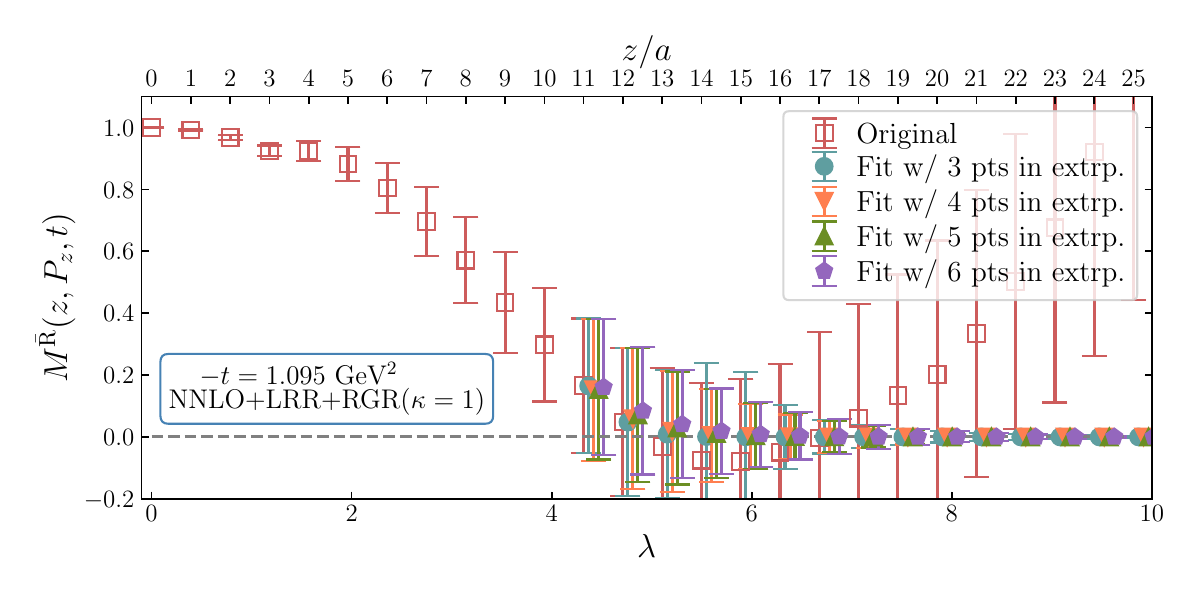}
    \caption{Comparison of extrapolated matrix elements obtained from different fit ranges.}
    \label{fig: extra}
\end{figure}

\begin{figure}[t!]
    \centering
    \includegraphics[width=0.6\linewidth]{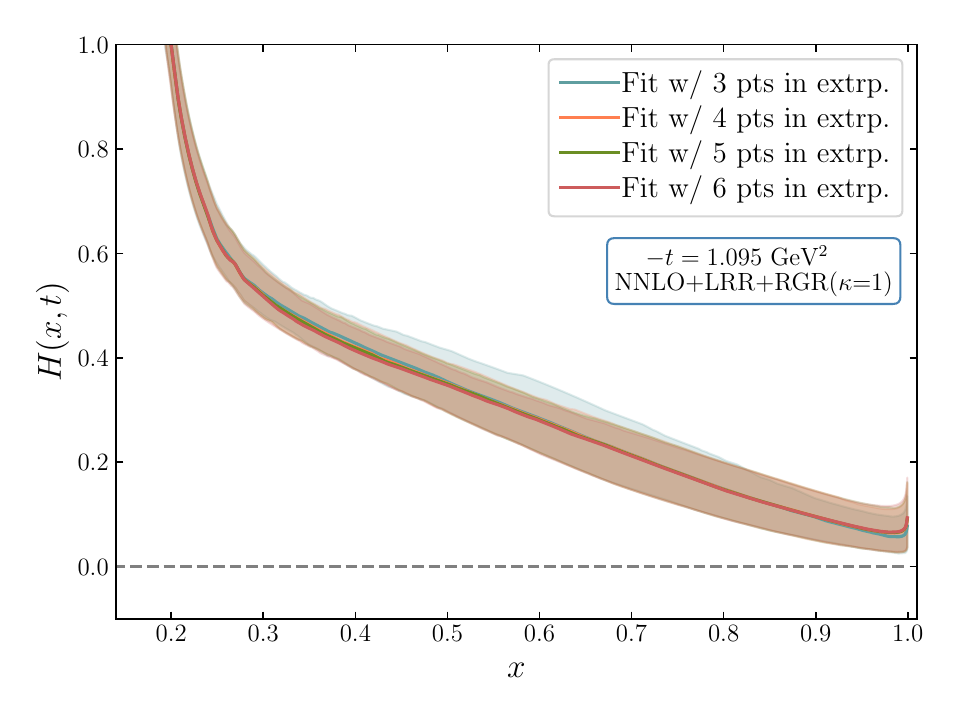} 
    \caption{Comparison of light-cone GPD results derived from various fit ranges in the extrapolation step.}
    \label{fig: compare GPD}
\end{figure}

To test the dependence of extrapolated matrix elements and light-cone GPD results on the fit range, we select four different values of $N\in[3,4,5,6]$, representing the number of data points used in the fit, to perform the fit with the 
decay model. Taking the case of $-t=1.095$ GeV$^2$ as an example, we compare renormalized matrix elements and light-cone GPD results in figure~\ref{fig: extra} and \ref{fig: compare GPD}, respectively. These results are all in good agreement, and the dependence on the fit range is very small. To better follow the extrapolation at long distances, we choose $N = 3$ for the analyses throughout this work.


\newpage
\section{\label{sec: t/pz2 dependence}Discussion of the $-t/P_z^2$ power correction}

\begin{figure}[t!]
    \centering	
    \vspace{-20pt}
    \includegraphics[width=0.6\linewidth]{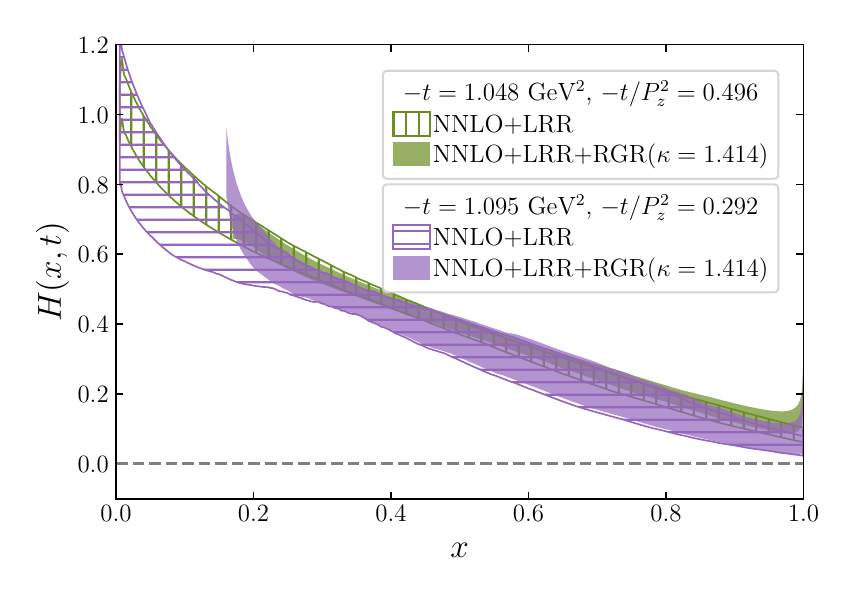}
    \caption{Comparison of LC GPDs with and without RGR.}
    \label{fig: t/pz2}
\end{figure}

In figure~\ref{fig: t/pz2}, we investigate potential power corrections by comparing two data sets with similar values of $-t$ but different $P_z$: $-t=1.048$ GeV$^2$ with $P_z=1.453$ GeV (yielding $-t/P_z^2\approx 0.45$ and $-t=1.095$ GeV$^2$ with $P_z=1.937$ GeV (yielding $-t/P_z^2\approx 0.29$). Despite the approximately 50\% difference in $-t/P_z^2$, the results obtained both with and without RGR are comparable between these two cases. This consistency is particularly relevant for our analysis at the largest value of $-t$ (1.69 GeV$^2$), which has a similar $-t/P_z^2\approx 0.45$, suggesting that the $-t/P_z^2$ power corrections are well under control even at our largest momentum transfer.

\section{\label{sec: matching} Perturbative coefficient}
\subsection{\label{subsec: matching} Matching coefficient}
The NNLO matching kernel with LRR and RGR can be expanded to ${\cal O}(\alpha_{s})$ as~\cite{Zhang:2023bxs, Holligan:2023rex}
\begin{equation}
	\begin{aligned}
		{\cal C}\left( \frac{x}{y}, \frac{\mu_0}{yP_z}, |y|\lambda_s \right) &= \delta\left(\frac{x}{y} - 1\right) + \frac{\alpha_{s}}{2\pi} {\cal C}^{(1)} \left( \frac{x}{y}, \frac{\mu_0}{yP_z}, |y|\lambda_s \right) + \left(\frac{\alpha_{s}}{2\pi}\right)^2 {\cal C}^{(2)} \left( \frac{x}{y}, \frac{\mu_0}{yP_z}, |y|\lambda_s \right) \\
		&\quad + \delta {\cal C}_{0, {\rm LRR\; rest}} {\cal C}_{\rm hybrid\; LRR}\left( \frac{x}{y}, \frac{\mu_0}{yP_z}, |y|\lambda_s \right) + {\cal O}(\alpha_{s}^3),
	\end{aligned}
\label{eq:matching}
\end{equation}
with
\begin{equation}
	\begin{aligned}
		\alpha_{s}(\mu_0) &= \frac{4 \pi a_{s0}}{X + a_{s0}\frac{\beta_1}{\beta_0} \ln(X) + \left(a_{s0}\right)^2Y}, \quad a_{s0} = \frac{0.293}{4 \pi}, \\
		X &= 1 + a_{s0} \beta_0 \ln\left(\left(\frac{\mu_0}{2}\right)^2\right), \\
		Y &= \frac{\beta_2}{\beta_0} \left(1 - \frac{1}{X}\right) + \left(\frac{\beta_1}{\beta_0}\right)^2 \left(\frac{\ln(X)}{X} + \frac{1}{X} - 1\right).
	\end{aligned}
\end{equation}
Here, the QCD $\beta$-function
\begin{equation}
    \beta[\alpha_{s}(\mu)] = - 2\alpha_s(\mu) \sum_{n=0}^\infty a_{s0}^{n+1}(\mu)\beta_n
\end{equation}
is given by
\begin{align}
\label{eq:beta}
\beta_0 &= {11\over 3}C_A - {4\over 3}T_Fn_f, \\
\beta_1 &= {34\over3}C_A^2 - \left({20\over3}C_A + 4C_F\right) T_Fn_f, \\
\beta_2 &= {2857\over 54}C_A^3 + \left(2C_F^2 - {205\over 9}C_FC_A - {1415\over 27}C_A^2 \right)T_Fn_f + \left({44\over 9} C_F+ {158\over 27}C_A \right)T_F^2 n_f^2,
\end{align}
where $C_F=4/3$, $C_A=3$, $T_F=1/2$, and $n_f=3$ for our lattice ensemble. 

The matching coefficient ${\cal C}$ can be derived from the NNLO kernel~\cite{Li:2020xml,Chen:2020ody} in the hybrid scheme~\cite{Ji:2020brr,Gao:2021dbh},
\begin{align}
    {\cal C}(\xi,{\mu_0\over p_z},z_sp_z) & = \left[{\cal C}_{\rm ratio}(\xi,{\mu_0\over p_z}) + {\cal C}_{\rm hybrid}(\xi,{\mu_0\over p_z},z_sp_z) \right]_+\,,
\end{align}
where $\xi=x/y$, $p_z=yP_z$, ${\cal C}_{\rm ratio}$ is the NNLO ratio scheme kernel~\cite{Orginos:2017kos,Izubuchi:2018srq}, and $\left[\ldots\right]_+$ denotes a plus function within the domain $-\infty<\xi<\infty$. The exact expressions for ${\cal C}_{\rm ratio}$ and ${\cal C}_{\rm hybrid}$ can be found in the attached Mathematica notebook.

In the second line of eq.~(\ref{eq:matching}), ${\cal C}_{\rm hybrid\; LRR}$ is the LRR term in the hybrid scheme~\cite{Zhang:2023bxs, Holligan:2023rex}
\begin{equation}
    \begin{aligned}
        &{\cal C}_{\rm hybrid\; LRR}(\xi,{\mu_0\over p_z},z_sp_z)\\
        &\equiv  \mu_0\int_{-\infty}^\infty {d(zp_z)\over 2\pi}e^{-i(1-\xi)zp_z} (|z|-z_s) e^{-\epsilon |z|}\theta (|z|-z_s)\\
        &= -{\mu_0\over p_z}\left\{ \frac{[(1-\xi)^2 -({\epsilon \over p_z})^2 ] \cos[(1-\xi) z_sp_z] + 2(1-x){\epsilon \over p_z}\sin[(1-\xi)z_sp_z]}{\pi [(1-\xi)^2+({\epsilon \over p_z})^2]^2}e^{-\epsilon z_s} \right\}_+, \\
    \end{aligned}
\end{equation}
The small parameter $\epsilon$ is introduced to make the Fourier transform converge, whose prescription differs from that in refs.~\cite{Zhang:2023bxs,Holligan:2023jqh} by $O(\epsilon)$, which has little impact on our final results. Moreover, the matching is insensitive to the value of $\epsilon$, and in our analysis we set $\epsilon=0.002a^{-1}$. Additionally, $\delta {\cal C}_{0, {\rm LRR\; rest}}$ represents the rest LRR term
\begin{align}
    \delta {\cal C}_{0, {\rm LRR\; rest}} &= {\cal C}_{0, {\rm LRR}} - \alpha_s {\cal C}^{(1)}_{0, {\rm LRR}}-\alpha_s^2 {\cal C}^{(2)}_{0, {\rm LRR}}\,,
\end{align}
where
\begin{equation}
\begin{aligned}
    {\cal C}_{0, {\rm LRR}} &= N_m \Bigg\lbrace (c_1+c_2) \alpha_s + {c_2(\beta_1-2\beta_0^2)\over 4\pi \beta_0}\alpha_s^2  \\
    &\quad - \frac{e^{-\frac{2 \pi}{\alpha_{s} \beta_0}} }{2 \beta_0^2} \left[ 4 \pi \beta_0 + \alpha_{s} c_1 \beta_1 + \frac{\alpha_{s}^2 \beta_1 c_2 (\beta_1 - 2 \beta_0^2)}{4 \pi \beta_0} \right] \Re\left[ E_{1 + \frac{\beta_1}{2 \beta_0^2}} \left(-\frac{2 \pi}{\alpha_{s} \beta_0}\right) \right]\Bigg\rbrace, \\
    {\cal C}^{(1)}_{0, {\rm LRR}} &= N_m (1+c_1+c_2),\\
    {\cal C}^{(2)}_{0, {\rm LRR}} &= N_m {2\beta_0^2(1- c_2) + \beta_1 (1+ c_1 + c_2) \over 4\pi \beta_0},\\
\end{aligned}
\end{equation}
with $E_n(z)=\int_{1}^{\infty}e^{-zt}/t^n {\rm d}t$ being the generalized exponential integral, $\Re$ denoting the real part, and 
\begin{equation}
\begin{aligned}
    c_1 &= \frac{\beta_1^2 - \beta_0\beta_2}{4b\beta_0^4},  \\
    c_2 &= \frac{\beta_1^4 + 4\beta_0^3\beta_1\beta_2 - 2\beta_0\beta_1^2\beta_2 + \beta_0^2(\beta_2^2-2\beta_1^3) - 2\beta_0^4\beta_3}{32 b (b - 1) \beta_0^8},
\end{aligned}
\end{equation}
with $b = \beta_1/(2\beta_0^2)$. Note that ${\cal C}^{(1)}_{0, {\rm LRR}}$ and ${\cal C}^{(2)}_{0, {\rm LRR}}$ are the lowest two terms in the $\alpha_s$ expansion of ${\cal C}_{0, {\rm LRR}}$, so $\delta {\cal C}_{0, {\rm LRR\; rest}}$ is ${\cal O}(\alpha_s^3)$ in perturbation theory.

Without LRR, the original hybrid matching kernel does not converge well~\cite{Zhang:2023bxs}. With LRR, the convergence of the matching kernel gets significantly improved, thanks to the subtraction of ${\cal C}^{(1)}_{0, {\rm LRR}}$ and ${\cal C}^{(2)}_{0, {\rm LRR}}$ terms,
\begin{align}\label{fullLRR}
    {\cal C} &= \delta\left(\frac{x}{y} - 1\right) + \frac{\alpha_{s}}{2\pi} \left[{\cal C}^{(1)} -(2\pi) {\cal C}^{(1)}_{0, {\rm LRR}} {\cal C}_{\rm hybrid\; LRR} \right] \nonumber \\
		&\qquad   + \left(\frac{\alpha_{s}}{2\pi}\right)^2 \left[{\cal C}^{(2)} -(2\pi)^2 {\cal C}^{(2)}_{0, {\rm LRR}}{\cal C}_{\rm hybrid\; LRR} \right] + {\cal C}_{0, {\rm LRR}} {\cal C}_{\rm hybrid\; LRR}.
\end{align}

In numerical implementation, we can discretize the matching kernel including the integration measure as a matrix,
\begin{align}\label{eq:mat}
    {dy\over |y|}{\cal C} \left({x\over y} \right) &= {\cal I}_{x,y} + \frac{\alpha_{s}}{2\pi} \bar{\cal C}^{(1)}_{x,y}  + \left(\frac{\alpha_{s}}{2\pi}\right)^2 \bar{\cal C}^{(2)}_{x,y} + {\cal C}^{{\rm LRR}}_{x,y}\,,
\end{align}
which corresponds to eq.~(\ref{fullLRR}) term by term.

To obtain the GPD from quasi-GPD, we need to first carry out the inverse matching. With a square matching matrix which is dominated by the diagonal elements, we can easily perform a matrix inversion. However, by doing so we lose the counting of the power of $\alpha_s$ or the perturbative accuracy, so it would be more natural to obtain the inverse kernel by expanding in $\alpha_s$. Nevertheless, a naive $\alpha_s$-expansion of the inverse of eq.~(\ref{eq:matching}) would not be convergent, despite that each term has a clear hierarchy in $\alpha_s$ order. Instead, we should use eq.~(\ref{eq:mat}) and treat ${\cal I} + {\cal C}^{{\rm LRR}} \equiv {\cal J}$ as the leading term, thus the inverse kernel is
\begin{align}\label{eq:inversemat}
    {\cal C}^{-1} &= {\cal J}^{-1} - \frac{\alpha_{s}}{2\pi}  {\cal J}^{-1}  \bar{\cal C}^{(1)}{\cal J}^{-1}  -  \left(\frac{\alpha_{s}}{2\pi}\right)^2 \left[ {\cal J}^{-1}  \bar{\cal C}^{(2)}{\cal J}^{-1} - {\cal J}^{-1}  \bar{\cal C}^{(1)}{\cal J}^{-1}\bar{\cal C}^{(1)}{\cal J}^{-1}\right]\,,
\end{align}
which would guarantee the convergence of perturbation theory and maintain the counting rule of perturbative accuracy.

At NLO+RGR+LRR accuracy, we truncate eq.~(\ref{eq:inversemat}) at ${\cal O}(\alpha_s)$ and use NLO DGLAP evolution and $\beta$-function. At NNLO+RGR+LRR accuracy, we truncate eq.~(\ref{eq:inversemat}) at ${\cal O}(\alpha_s^2)$ and use NNLO DGLAP evolution and $\beta$-function.

\subsection{\label{subsec: evolution} Evolution coefficient}
The NNLL matching requires the 3-loop DGLAP evolution kernel, which can be expressed as
\begin{equation}
\begin{aligned}
    {\cal C}_{\rm evo}\left( a_s, \frac{x}{y}, \frac{\mu_0}{\mu} \right) = \;&\delta\left( \frac{x}{y} - 1 \right) + a_s t {\cal P}^{(0)}_{qq} \left( \frac{x}{y} \right) \\
    & + a_s^2  \left[ t{\cal P}^{V(1)}_{qq} + {t^2\over 2}\left( {\cal P}^{(0)}_{qq} \otimes {\cal P}^{(0)}_{qq} + \beta_0 {\cal P}^{(0)}_{qq} \right)\right] \left( \frac{x}{y} \right) \\
    & + a_s^3 \left[t{\cal P}_{qq}^{V(2)} + t^2\left({\beta_1\over2}{\cal P}_{qq}^{(0)} + {\beta_0}{\cal P}_{qq}^{V(1)} + {\cal P}_{qq}^{(0)}\otimes {\cal P}_{qq}^{V(1)}\right) \right.\\
    & +\left.{t^3\over 6}\!\left(2\beta_0^2 {\cal P}_{qq}^{(0)} \!+\! 3\beta_0{\cal P}_{qq}^{(0)}\!\otimes\! {\cal P}_{qq}^{(0)} \!+\! {\cal P}_{qq}^{(0)}\!\otimes\! {\cal P}_{qq}^{(0)}\!\otimes\! {\cal P}_{qq}^{(0)}\right)\!\right]\!\left( \frac{x}{y} \right),
\end{aligned}
\end{equation}
where $a_s=\alpha_{s}(\mu_0) / (4\pi)$, $t= \ln(\mu^2_0/\mu^2)$, ${\cal P}^{(0)}_{qq}$ is the 1-loop splitting function, and ${\cal P}^{V(1)}_{qq}$ and ${\cal P}^{V(2)}_{qq}$ are the 2-loop and 3-loop splitting functions for the valence quark~\cite{Curci:1980uw, Moch:2004pa}. They are given by
\begin{equation}
\begin{aligned}
    {\cal P}^{(0)}_{qq} \left( z \right) &= 2C_F\left[\frac{1+z^2}{1-z}\right]_+, \\
    {\cal P}^{V(1)}_{qq} \left( z \right) &= 4\left[{\cal P}^{(1)}_{qq}(z) - {\cal P}^{(1)}_{q\bar{q}}(z)\right],\\
    {\cal P}^{(1)}_{qq}(z) &= C_F\left[C_F P_F(z)+\frac{1}{2}C_A P_G(z)+n_fT_F P_{n_f}(z)\right],\\
    {\cal P}^{(1)}_{q\bar{q}}(z) &= C_F\left(C_F-{1\over2}C_A\right) P_A(z),
\end{aligned}
\end{equation}
where the definitions of $P_F(z)$, $P_G(z)$, $P_{n_f}(z)$ and $P_A(z)$ can be found in ref.~\cite{Curci:1980uw}. As for ${\cal P}^{V(2)}_{qq}$, we take the approximate solution ${\cal P}^{V(2)}_{qq}(z)= P_{\rm ns}^{(2)-}(z)$ in eq.~(4.23) of ref.~\cite{Moch:2004pa}.

Note that after the inverse matching, one obtains the matched GPDs $H(x, \mu_0=2\kappa xP_z)$ at a varying $\overline{\rm MS}$ scale in $x$. To implement the DGLAP evolution, we set $\mu=2$ GeV and $\mu_0=2\kappa xP_z$ and obtain the evolution kernel as a triangular matrix down to a minimal value $x_{\rm min}\sim 0.15$ for $P_z=1.937$ GeV. Then we invert the triangular matrix and apply it to the matched GPDs $H(x, \mu_0)$, which leads to $H(x,\mu)$ at the fixed scale $\mu$ down to $x_{\rm min}\sim 0.15$.

We could also express the evolution kernel ${\cal C}_{\rm evo}$ as a perturbative series in $a_s(\mu)$, but it gives a smaller scale variation as $a_s$ remains small. Instead, we use ${\cal C}_{\rm evo}$ as a perturbative series in $a_s(2\kappa xP_z)$. The latter becomes large at small $x$, which gives us a more conservative estimate of the scale variation uncertainty.

\bibliographystyle{JHEP}
\bibliography{ref.bib}

\end{document}